\documentclass[final,3p,times,twocolumn]{elsarticle}
\usepackage{amssymb,amsmath,amsthm,bm}
\usepackage{geometry}
\usepackage{algorithm}
\usepackage{algpseudocode}

\usepackage[retainorgcmds]{IEEEtrantools}

\usepackage{graphicx}
\graphicspath{{figures-pdf/}}

\usepackage[normalem]{ulem}
\usepackage{url}
\usepackage{lineno}
\usepackage{bm}
\usepackage{graphicx,color,overpic}

\newlength\imagewidth
\newlength\figwidth
\newlength\sfigwidth
\newlength\vfigskip
\journal{Elsevier}
\usepackage{color}
\definecolor{dgreen}{rgb}{0,.6,0}
\newtheorem{definition}{Definition}

\usepackage{hyperref}
\hypersetup{
    colorlinks=true,
    linkcolor=blue,
    filecolor=magenta,
    urlcolor=cyan,
}

\begin{document}

\begin{frontmatter}

\title{Cracking a hierarchical chaotic image encryption algorithm based on permutation}
%\thanks{A preliminary version of this paper .}

\author{Chengqing Li\corref{corr}}
\ead{DrChengqingLi@gmail.com}
% [cn-xtu-cie]

\cortext[corr]{Corresponding author.}

%MOE (Ministry of Education) Key Laboratory of Intelligent Computing and Information Processing,\\
\address{College of Information Engineering, Xiangtan University, Xiangtan 411105, Hunan, China}
%\address[cn-xtu-icip]{, Xiangtan University, China}

\begin{abstract}
In year 2000, an efficient hierarchical chaotic image encryption (HCIE) algorithm was proposed, which divides a plain-image
of size $M\times N$ with $T$ possible value levels into $K$ blocks of the same size and then operates position permutation on two levels: intra-block and inter-block. As a typical position permutation-only encryption algorithm, it has received intensive attention. The present paper analyzes specific security performance of HCIE against ciphertext-only attack and known/chosen-plaintext attack. It is found that only $O(\lceil\log_T(M\cdot N/K) \rceil)$ known/chosen plain-images are sufficient to achieve a good performance, and
the computational complexity is $O(M\cdot N\cdot \lceil\log_T(M\cdot N/K) \rceil)$, which effectively demonstrates that hierarchical permutation-only image
encryption algorithms are less secure than normal (i.e., non-hierarchical) ones. Detailed experiment results are given to verify the feasibility of the known-plaintext attack. In addition, it is pointed out that the security of HCIE against ciphertext-only attack was much overestimated.
\end{abstract}
\end{frontmatter}

\section{Introduction}

With the increasing transmission speeds of wired/wireless networks and popularization of image capturing devices and cloud storage
services, image data are transmitted over open networks more and more frequently. This makes security of image data become
more and more important. The public concern of it becomes serious as news about the illegal online leak of personal photos of some celebrities was released.
As a chaotic system owns some similar properties as that of modern encryption schemes, it has been intensively studied as an alternative approach for designing
secure and efficient encryption schemes \cite{Li:Dissertation2003,YaobinMao:CSF2004,AlvarezLi:Rules:IJBC2006}. The main idea and principle of applying chaos theory to protecting images can be traced back to 1986 \cite{Crutchfield:Chaos:SA86}, which demonstrates the stretching effect of a chaotic map on a painting of Henri Poincar\'{e}, a founder of modern chaos theory.

The simplest and most efficient method for protecting multimedia data is permuting the positions of their spatial pixels \cite{Matias&Shamir:CurveImageEncryption:Crypto87} or frequency coefficients \cite{Zeng:VideoScrambling:IEEETMM2003}. In the literature, some synonyms of permutation, transposition, shuffle, scramble \cite{Zeng:VideoScrambling:IEEETMM2003}, swap and shift,
are used. Security scrutiny on some specific permutation-only encryption algorithms against known/chosen-plaintext attacks were previously developed \cite{Jan-Tseng:BreakingSCAN:IPL96,Chang-Yu:BreakingSCAN:PRL2002}. In \cite{WHLi:shuffle:ACMM2012}, a ciphertext-only attack on
a specially simple permutation-only encryption algorithm was proposed utilizing correlation redundancy remaining in the cipher-image. No matter how the permutation relationships are generated and what the permutation object is,
any permutation-only encryption algorithm can always be represented by a \textit{permutation relationship matrix}, whose entry
stores the corresponding permuted location in the cipher-text \cite{Li:Permutation:SPIC2008}. The security of permutation-only encryption algorithm
relies on its \textit{real permutation domain}, in which any element in the permutation object can be permuted independently. As for a permutation domain of size $M\times N$ with $T$ possible value levels, it is estimated that the required number of known/chosen-plaintexts for an efficient plaintext attack is $O(\lceil\log_T(M\cdot N)\rceil)$, where $\lceil x\rceil$ denotes the ceiling function. An upper bound of the attack complexity is also derived therein to be
$O(n\cdot (M\cdot N)^2)$, where $n$ is the number of known/chosen plain-images \cite{Li:Permutation:SPIC2008}. In \cite{Lcq:Optimal:SP11}, the computational complexity of the attack is
further reduced to $O(n\cdot (M\cdot N))$ by replacing the set intersection operations of quadratic complexity with linear element access operations.
Even so, all kinds of permutation operations are still being used in multimedia protection today \cite{Li:scramble:ITCSVT2008,Sohn:scrambling:TCSVT2011,Zhouyc:perturbation:SP14,SMYU:ARM:IJBC2014}.

In \cite{Yen-Guo:HCIE:IEEPVISP2000}, a typical example of permutation-only image encryption algorithms, called HCIE (hierarchical chaotic image encryption),
was proposed. Although security performance of general permutation-only image encryption algorithms against plaintext attack has been
quantitatively analyzed, specific security performance of HCIE is still not evaluated. The core of HCIE is a
permutation function composed of rotation operations of four directions, originates from an intellectual toy, Rubik's Cube \cite{Korf:Cube:97}.
In \cite{Yen-Guo:HCIE:IEEPVISP2000}, the authors claimed about the security property of HCIE as follows: ``By way of collecting some original images and their
encryption results or collecting some specified images and their corresponding encryption results, it is still difficult for
the cryptanalysts to decrypt an encrypted image correctly because the permutation relationship is different for each
image." In this paper, we will demonstrate that the claim on the robustness of HCIE against known/chosen-plaintext attack is groundless. Further more, we find that the hierarchical encryption structure suggested in HCIE does not provide any higher security against known/chosen-plaintext attack, but actually make
the overall security even weaker. In addition, we find the capability of HCIE against ciphertext-only attack was much over-estimated.

The rest of this paper is organized as follows. The algorithm HCIE is briefly introduced in Sec.~\ref{sec:HCIE}.
Detailed cryptanalysis on HCIE is provided in Sec.~\ref{sec:cryptanalysisHCIE2}, with some experimental results.
The last section concludes the paper.

\section{The hierarchical chaotic image encryption algorithm (HCIE)}
\label{sec:HCIE}

HCIE is a two-level hierarchical permutation-only image encryption algorithm, in which all involved permutation relationships are defined by pseudo-random
combinations of four rotation mappings with pseudo-random
parameters. For an image, $f=[f(i,j)]_{M\times N}$, the four
mapping operations are described as follows, where $p<\min(M,N)$
holds for each mapping.

\begin{definition}
The mapping $f'=ROLR^{i,p}_b(f)$ $(0\leq i\leq M-1)$ is defined to
rotate the $i$-th row of $f$, in the left (when $b=0$) or right
(when $b=1$) direction, by $p$ pixels.
\end{definition}

\begin{definition}
The mapping $f'=ROUD^{j,p}_b(f)$ $(0\leq j\leq N-1)$ is defined to
rotate the $j$-th column of $f$, in the up (when $b=0$) or down
(when $b=1$) direction, by $p$ pixels.
\end{definition}

\begin{definition}
The mapping $f'=ROUR^{k,p}_b(f)$ $(0 \leq k \leq M+N-2)$ is
defined to rotate all pixels satisfying $i+j=k$, in the lower-left
(when $b=0$) or upper-right (when $b=1$) direction, by $p$ pixels.
\end{definition}

\begin{definition}
The mapping $f'=ROUL^{l,p}_b(f)$ $(1-N \leq l \leq M-1)$ is
defined to rotate all pixels satisfying $i-j=l$, in the upper-left
(when $b=0$) or lower-right (when $b=1$) direction, by $p$ pixels.
\end{definition}

Given a pseudo-random bit sequence $\{b(i)\}$ starting from $i_0$,
the \texttt{Sub\_HCIE} function in Algorithm~1 is used to permute an
$S_M\times S_N$ image $f_{sub}$ to become another $S_M\times S_N$
image $f'_{sub}$, where $(\alpha,\beta,\gamma,no)$ are control
parameters.
\begin{algorithm}
\caption{The \texttt{Sub\_HCIE} function}
\begin{algorithmic}[1]
\Function {\texttt{Sub\_HCIE}}{$f_{sub}$, $\{b(i)\}$, $no$, $S_M$, $S_N$}
\For {$ite \leftarrow 0, no$}
  \State $q \leftarrow i_0+(3S_M+3S_N-2)\cdot ite$
  \State $p \leftarrow \alpha+\beta \cdot b(q+0)+\gamma \cdot b(q+1)$
  \For {$i\leftarrow 0, (S_M-1)$}
  \State $f'_{sub}\leftarrow ROLR^{i,p}_{b(i+q)}(f_{sub})$
  \EndFor
  \For {$j\leftarrow 0, (S_N-1)$ }
  \State $f'_{sub}\leftarrow ROUD^{j,p}_{b(j+q+S_M)}(f'_{sub})$
  \EndFor
  \For {$k\leftarrow 0, (S_M+S_N-2)$}
  \State $f'_{sub}\leftarrow ROUR^{k,p}_{b(k+q+S_M+S_N)}(f'_{sub})$
  \EndFor
  \For {$l\leftarrow (1-S_N), (S_M-1)$}
  \State $f'_{sub}\leftarrow ROUL^{l,p}_{b(l+q+2\cdot S_M+3\cdot S_N-2)}(f'_{sub})$
  \EndFor
\EndFor
\State $i_0\leftarrow i_0+(3S_M+3S_N-2)\cdot no$
\State \textbf{return} $(f'_{sub}, i_0)$
\EndFunction
\end{algorithmic}
\end{algorithm}
One can see that the \texttt{Sub\_HCIE} function actually
defines an $S_M\times S_N$ permutation relationship matrix pseudo-randomly
controlled by $(3S_M+3S_N-2)\cdot no$ bits in the bit sequence
$\{b(i)\}$ from $i_0$. Based on this function, for an $M\times N$
image $f=[f(i,j)]_{M\times N}$, the four basic parts of HCIE
can be briefly described as follows.
\begin{itemize}
\item \textit{The secret key} is the initial condition $x(0)$ and
the control parameter $\mu$ of the chaotic Logistic map, $f(x)=\mu
x(1-x)$ \cite{Devaney:Chaos:2003}, which is realized in $L$-bit finite precision.

\item \textit{Some public parameters}: $S_M$, $S_N$, $\alpha$,
$\beta$, $\gamma$ and $no$, where $\sqrt{M}\leq S_M \leq M$,
$M\bmod S_M=0$, $\sqrt{N}\leq S_N\leq N$, and $N\bmod S_N=0$.

Note: {\it Although ($S_M,S_N,\alpha,\beta,\gamma,no$) can all be
included in the secret key, they are not suitable for such a use
due to the following reasons: 1) $S_M,S_N$ are related to $M,N$;
2) $\alpha,\beta,\gamma$ are related to $S_M,S_N$ (and then
related to $M,N$, too); 3) $S_M,S_N$ can be easily guessed from
the mosaic effect of the cipher-image; 4) iteration number $no$ cannot be too large
to achieve an acceptable encryption speed.}

\item \textit{The initialization procedure} of generating the bit
sequence used in the \texttt{Sub\_HCIE} function: run the Logistic
map starting from $x(0)$ to generate a chaotic sequence,
$\{x(i)\}_{i=0}^{\lceil L_b/8\rceil-1}$, and then extract 8
bits following the decimal point of each chaotic state $x(i)$ to
yield a bit sequence $\{b(i)\}_{i=0}^{L_b-1}$, where
$L_b=\left(1+\frac{M}{S_M}\cdot\frac{N}{S_N}\right)\cdot(3S_M+3S_N-2)\cdot
no$; finally, set $i_0=0$ to let the \texttt{Sub\_HCIE} function
run starting from $b(0)$.

\item \textit{The two-level hierarchical encryption procedure}:

\textit{1) The high-level encryption -- permuting image
blocks}: divide the plain-image $f$ into blocks of size $S_M\times
S_N$, which compose an $\frac{M}{S_M}\times\frac{N}{S_N}$
block-image
\[P_f=\left[P_f(i,j)\right]_{\frac{M}{S_M}\times\frac{N}{S_N}},\]
where $P_f(i,j)$ is the block of size $S_M\times S_N$ at the
position $(i,j)$. Then, permute the positions of all blocks with
the \texttt{Sub\_HCIE} function in the following way:
a) create a pseudo-image $f_p=[f_p(i,j)]_{S_M\times S_N}$
containing $\left(\frac{M}{S_M}\cdot\frac{N}{S_N}\right)$ non-zero
indices of all image blocks in $P_f$ and $\left(S_M\cdot
S_N-\frac{M}{S_M}\cdot\frac{N}{S_N}\right)$ zero-elements, and
permute $f_p$ with the \texttt{Sub\_HCIE} function to get a
shuffled pseudo-image $f_p^*$; b) generate a permuted block-image $P_{f^*}$ from $P_f$ (i.e.,
permute $f$ blockwise) using the shuffled indices contained in
$f_p^*$.
The above high-level encryption procedure can be considered as a
permutation of the block-image:
$P_f\xrightarrow{f_p^*=\mathtt{Sub\_HCIE}(f_p)}P_{f^*}$, where
$f_p^*$ actually corresponds to an
$\frac{M}{S_M}\times\frac{N}{S_N}$ permutation relationship matrix.

\textit{2) The low-level encryption -- permuting pixels in every
image block one by one}: for $i=0\sim\left(\frac{M}{S_M}-1\right)$ and
$j=0\sim\left(\frac{N}{S_N}-1\right)$, call the \texttt{Sub\_HCIE}
function to permute each block $P_{f^*}(i,j)$ so as to get the
corresponding block of the cipher-image $f'$:
\[P_{f'}(i,j)=\mathtt{Sub\_HCIE}\left(P_{f^*}(i,j)\right).
\]
\end{itemize}

As normalized in \cite{Li:Permutation:SPIC2008}, any permutation-only encryption algorithm exerting on
an object of size $H\times W$ can be represented with a \textit{permutation relationship matrix} of size
$H\times W$, denoted by
\begin{equation}
\bm{W}=\left[w(i,j)=(i',j')\in\mathbb{H}\times\mathbb{W}\right]_{H\times W},
\label{eq:PermutationRelation}
\end{equation}
where $\mathbb{H}=\{0,\cdots, H-1\}$,
$\mathbb{W}=\{0,\cdots, W-1\}$, and
$w(i_1,j_1)\neq w(i_2,j_2)$ for any $(i_1,j_1)\neq(i_2,j_2)$.

In HCIE, a total of
$\left(1+\frac{M}{S_M}\cdot\frac{N}{S_N}\right)$ permutation relationship
matrices are involved: 1) one high-level permutation relationship matrix of
size $\frac{M}{S_M}\times\frac{N}{S_N}$; 2)
$\left(\frac{M}{S_M}\cdot\frac{N}{S_N}\right)$ low-level
permutation relationship matrices of size $S_M\times S_N$. With the
above-mentioned representation of a permutation-only image encryption algorithm,
the secret key $(\mu,x(0))$ of HCIE is equivalent to the
$\left(1+\frac{M}{S_M}\cdot\frac{N}{S_N}\right)$ permutation relationship matrices
for plain-images of the same size. To facilitate the following discussions, we use
$\bm{W}_0=[w_0(i,j)]_{\frac{M}{S_M}\times\frac{N}{S_N}}$ to denote
the high-level permutation relationship matrix, and use
$\left\{\bm{W}_{(i,j)}\right\}_{i=0,j=0}^{\frac{M}{S_M}-1,\frac{N}{S_N}-1}$
to denote the $\left(\frac{M}{S_M}\times\frac{N}{S_N}\right)$
low-level permutation relationship matrices, where
$\bm{W}_{(i,j)}=\left[w_{(i,j)}(i',j')\right]_{S_M\times S_N}$.
Apparently, the $\left(1+\frac{M}{S_M}\cdot\frac{N}{S_N}\right)$
permutation relationship matrices can be easily transformed to an equivalent
permutation relationship matrix of size $M\times N$, $\bm{W}=[w(i,j)]_{M\times
N}$.

When $S_M=M$ and $S_N=N$ (or $S_M=S_N=1$), the two hierarchical
encryption levels merge into a single one; the
$\left(1+\frac{M}{S_M}\cdot\frac{N}{S_N}\right)$ permutation relationship matrices become one permutation relationship matrix of size $M\times N$,
a typical permutation-only image encryption algorithm in which each pixel can be independently
permuted to any other position in the whole image by a single
$M\times N$ permutation relationship matrix $\bm{W}$.

\section{Cryptanalysis of HCIE}
\label{sec:cryptanalysisHCIE2}

\subsection{Ciphertext-only attack}

In \cite{Yen-Guo:HCIE:IEEPVISP2000}, it was claimed that the complexity of
brute-force attacks to HCIE is $O\left(2^{L_b}\right)$, since
there are
$L_b=\left(1+\frac{M}{S_M}\cdot\frac{N}{S_N}\right)\cdot(3S_M+3S_N-2)\cdot
no$ secret chaotic bits in $\{b(i)\}_{i=0}^{L_b-1}$ that are
unknown to the attackers. However, this statement is not true due to
the following fact: the $L_b$ bits are uniquely determined by the
secret key, i.e., the initial condition $x(0)$ and the control
parameter $\mu$, which have only $2L$ secret bits. This means that
there are only $2^{2L}$ different chaotic bit sequences.

Now, let
us study the real complexity of brute-force attacks. For each pair
of guessed values of $x(0)$ and $\mu$, the following operations
are needed:
\begin{itemize}
\item generating the chaotic bit sequence: there are $L_b/8$ chaotic
iterations;

\item creating the pseudo-image $f_p$: the complexity is $S_M\cdot
S_N$;

\item shuffling the pseudo-image $f_p$: running the
\texttt{Sub\_HCIE} function once;

\item generating $P_{f^*}$: the complexity is $M\cdot N$;

\item shuffling the partition image $P_{f^*}$: running the
\texttt{Sub\_HCIE} function for
$\left(\frac{M}{S_M}\cdot\frac{N}{S_N}\right)$ times.
\end{itemize}
Assume that the computing complexity of the \texttt{Sub\_HCIE}
function is $(4S_M+4S_N)\cdot no$. Then, the total complexity of
brute-force attacks to HCIE can be estimated to be
$O\left(2^{2L}\cdot(L_b+M\cdot N)\right)$, which is much smaller than
$O\left(2^{L_b/8}\right)$ when $L_b$ is not too small.
Additionally, considering the fact that the Logistic map can
exhibit a sufficiently strong chaotic behavior only when $\mu$ is
close to 4 \cite{Li:logistic:ND2014}, the above complexity should
be even smaller. This analysis shows that the security of
HCIE was much over-estimated by the authors in
\cite{Yen-Guo:HCIE:ISC99,Yen-Guo:HCIE:IEEPVISP2000}, for brute-force attacks.

Observing the hierarchical permutation structure of HCIE, one can see that
the histogram of each $S_M\times S_N$ block in the plain-image will keep unchanged
during the whole permutation process of HCIE. Due to the strong correlation between neighbouring pixels
(and even blocks) of natural images (See \cite[Fig.~5]{Li:RCES:JSS2008}), there exists some correlation between histograms of neighbouring blocks.
So, one may determine the relative locations of some blocks in cipher-image by comparing similarity degrees of
histograms for every pair of cipher-blocks \cite{HSLI:Banknote:TM14,Ling:histogramcomparison:PAMI07}.

\subsection{The known-plaintext attack}

Since HCIE is a permutation-only image encryption algorithm, given $n$ known
plain-images $f_1\sim f_n$ of size $M\times N$ and the
corresponding cipher-images $f'_1\sim f'_n$, one can simply call
the \texttt{Get\_Permutation\_Matrix} function defined in \cite[Sec.~3.1]{Li:Permutation:SPIC2008} or its enhanced version in
\cite[Sec.~4]{Lcq:Optimal:SP11} with the input parameter $(f_1\sim f_n,f'_1\sim f'_n, M, N)$ to estimate an
$M\times N$ permutation relationship matrix $\bm{W}$, which is equivalent to
the $\left(1+\frac{M}{S_M}\cdot\frac{N}{S_N}\right)$ smaller
permutation relationship matrices. However, if the hierarchical structure of
HCIE is considered, the known-plaintext attack may be quicker and
the estimation will be more effective. Thus, the following hierarchical procedure of
known-plaintext attacks to HCIE is suggested\footnote{For HCIE,
the permutation relationship matrices also depend on the values of the public
parameters. To simplify the following description, without loss of
generality, it is assumed that all public parameters are fixed for
all known plain-images.}:
\begin{itemize}
\item \textit{Reconstruct the high-level permutation relationship matrix
$\bm{W}_0$}: \textit{1)  for $i=0\sim\left(\frac{M}{S_M}-1\right)$ and
$j=0\sim\left(\frac{N}{S_N}-1\right)$}: calculate the mean values of the $2n$ blocks $P_{f_1}(i,j)\sim P_{f_n}(i,j)$,
$P_{f'_1}(i,j)\sim P_{f'_n}(i,j)$ and denote them by
$\overline{P_{f_1}(i,j)}\sim \overline{P_{f_n}(i,j)}$ and
$\overline{P_{f'_1}(i,j)}\sim \overline{P_{f'_n}(i,j)}$;

\textit{2) generate $2n$ images $\overline{P}_{f_1}\sim
\overline{P}_{f_n}$ and $\overline{P}_{f'_1}\sim
\overline{P}_{f'_n}$ of size $\frac{M}{S_M}\times\frac{N}{S_N}$ as
follows}: $\forall m=1\sim n$,
\begin{equation}
\overline{P}_{f_m}=\left[\overline{P_{f_m}(i,j)}\right]_{\frac{M}{S_M}\times\frac{N}{S_N}}
\label{equation:PlainBlock}
\end{equation}
and
\begin{equation}
\overline{P}_{f_m'}=\left[\overline{P_{f_m'}(i,j)}\right]_{\frac{M}{S_M}\times\frac{N}{S_N}},
\label{equation:CipherBlock}
\end{equation}
and call the \texttt{Get\_Permutation\_Matrix} function with the
input parameters
\[
\left(\overline{P}_{f_1}\sim \overline{P}_{f_n},
\overline{P}_{f'_1}\sim \overline{P}_{f'_n}, \frac{M}{S_M},
\frac{N}{S_N}\right)
\]
to get an estimated permutation relationship matrix
$\widetilde{\bm{W}}_0=\left[\widetilde{w}_0(i,j)\right]_{\frac{M}{S_M}\times\frac{N}{S_N}}$
and its inverse
$\widetilde{\bm{W}}_0^{-1}=\left[\widetilde{w}_0^{-1}(i,j)\right]_{\frac{M}{S_M}\times\frac{N}{S_N}}$.
\end{itemize}

\textit{3) \textit{Reconstruct the
$\left(\frac{M}{S_M}\cdot\frac{N}{S_N}\right)$ low-level
permutation relationship matrices
$\left\{\bm{W}_{(i,j)}\right\}_{i=0,j=0}^{\frac{M}{S_M}-1,\frac{N}{S_N}-1}$}}:
\begin{itemize}
\item for $i=0\sim\left(\frac{M}{S_M}-1\right)$ and
$j=0\sim\left(\frac{N}{S_N}-1\right)$, call the
\texttt{Get\_Permutation\_Matrix} function with the input
parameters $(P_{f_1}(i,j)\sim P_{f_n}(i,j), P_{f'_1}(i',j')\sim
P_{f'_n}(i',j'), S_M, S_N)$, where $(i',j')=W_0(i,j)$, to
determine an estimated permutation relationship matrix
$\widetilde{\bm{W}}_{(i,j)}$ and its inverse
$\widetilde{\bm{W}}_{(i,j)}^{-1}$.
\end{itemize}

With the $\left(1+\frac{M}{S_M}\cdot\frac{N}{S_N}\right)$ inverse
matrices $\bm{W}_0^{-1}$ and
$\left\{\widetilde{\bm{W}}_{(i,j)}^{-1}\right\}_{i=0,j=0}^{\frac{M}{S_M}-1,\frac{N}{S_N}-1}$,
one can decrypt a new cipher-image $f_{n+1}'$ with \texttt{Dermutation} function
given in Algorithm~2 to get an estimated plain-image $f_{n+1}^*$:
\begin{algorithm}
\caption{The function \texttt{Dermutation}}
\begin{algorithmic}[1]
\Function{\texttt{Dermutation}}{$\bm{W}_0^{-1}$, $\left\{\widetilde{\bm{W}}_{(i,j)}^{-1}\right\}_{i=0,j=0}^{\frac{M}{S_M}-1,\frac{N}{S_N}-1}$, $f_{n+1}'$}
\For{$i \leftarrow 0, (M/S_M)-1$}
  \For {$j \leftarrow 0, (N/S_N)-1$}
     \State $f_{temp} \leftarrow P_{f'_{n+1}}(w_0^{-1}(i,j))$
        \For{$ii\leftarrow 0, S_M-1$}
           \For{ $jj\leftarrow 0, S_N-1$ }
              \State $f_{temp}^*(ii,jj)\leftarrow f_{temp}\left(w_{(i,j)}^{-1}(ii,jj)\right)$
              \State $P_{f_{n+1}^*}(i,j)\leftarrow f_{temp}^*$
            \EndFor
      \EndFor
  \EndFor
\EndFor
\State \textbf{return} $f_{n+1}^*$
\EndFunction
\end{algorithmic}
\end{algorithm}

In fact, in the above procedure, any measure keeping invariant in
the block permutations can be used instead of the mean value. A
typical measure is the histogram of each $S_M\times S_N$ block.
Although the mean value is less precise than the histogram, it
works well in most cases and is effective to reduce the time
complexity. When $T$ and $S_M\times S_N$ are both very small, the
efficiency of the mean value will become low, in this case the histogram or
the array of all pixel values can be used as a replacement.
As for an image of size $H\times W$ and $T$ possible value levels,
the number of different histogram is
\begin{equation}
n_h=\sum_{i=1}^{\min(T,\ H\cdot W)} {T\choose i}\cdot {H\cdot W\choose i-1}.
\end{equation}
As $n_h$ is a huge number for a non-tiny image and histogram is sensitive to the change of pixel value, it is easier to get the high-level
permutation relationship matrix $\bm{W}_0$ than the low-level permutation relationship matrices.

Finally, let us see whether the hierarchical structure used in
HCIE is helpful for enhancing the security against the
known-plaintext attack to the common permutation image ciphers. As
discussed in \cite[Sec.~3.1]{Li:Permutation:SPIC2008}, $n\geq\lceil\log_T(2(M\cdot N-1))\rceil$ known
plain-images are needed to achieve an acceptable breaking
performance. Since the hierarchical structure makes it possible
for an attacker to work on permutation relationship matrices of size $S_M\times
S_N$ or $\frac{M}{S_M}\times\frac{N}{S_N}$ (both smaller than
$M\times N$), it is obvious that for HCIE the number of required
known plain-images will be smaller than
$\lceil\log_T(2(M\cdot N-1))\rceil$. As the permutation relationship matrix $\bm{W}_0$
can be recovered in a very high probability with one or two known plain-images, one
can conclude as follows: the smaller the $(S_M\times S_N)$ is, the smaller the $n$ is. Also, the attack complexity will
become lower, since the number of used plaintexts is reduced. In this sense, hierarchical permutation-only image
ciphers are less secure than non-hierarchical ones, which discourages the use of HCIE.

\subsection{Experimental results on known-plaintext attack}
\label{section:Experiments}

\begin{figure}[!htb]
\centering
\begin{minipage}{\sfigwidth}
\centering
\includegraphics[width=\textwidth]{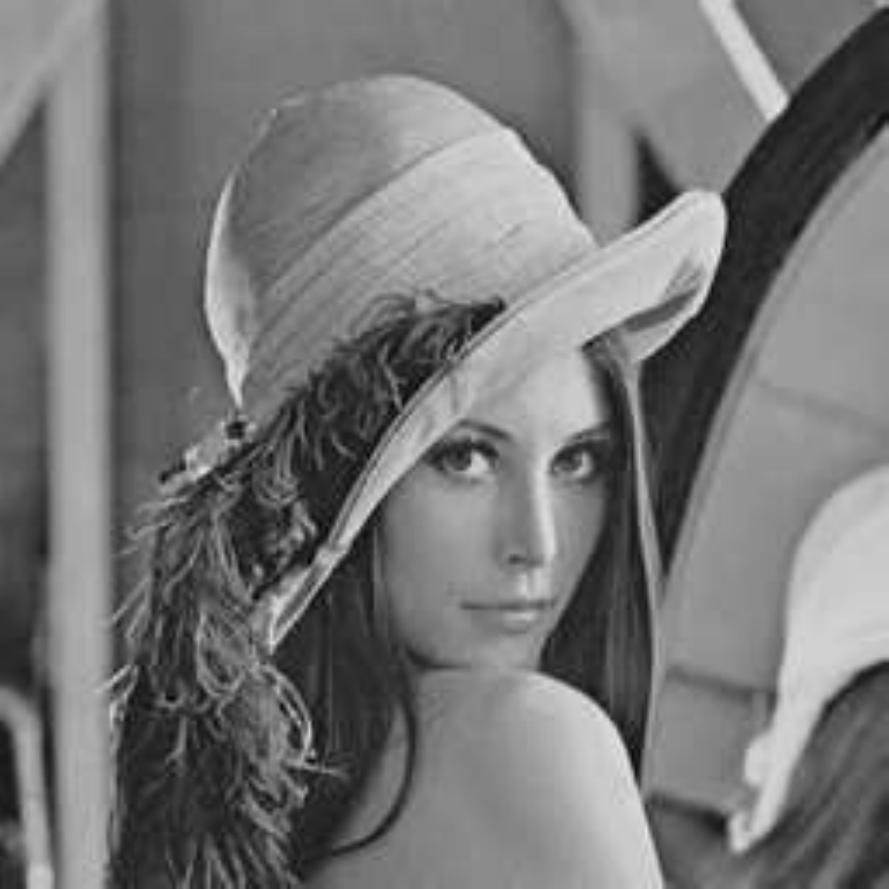}\\
Image \#1
\end{minipage}
\begin{minipage}{\sfigwidth}
\centering
\includegraphics[width=\textwidth]{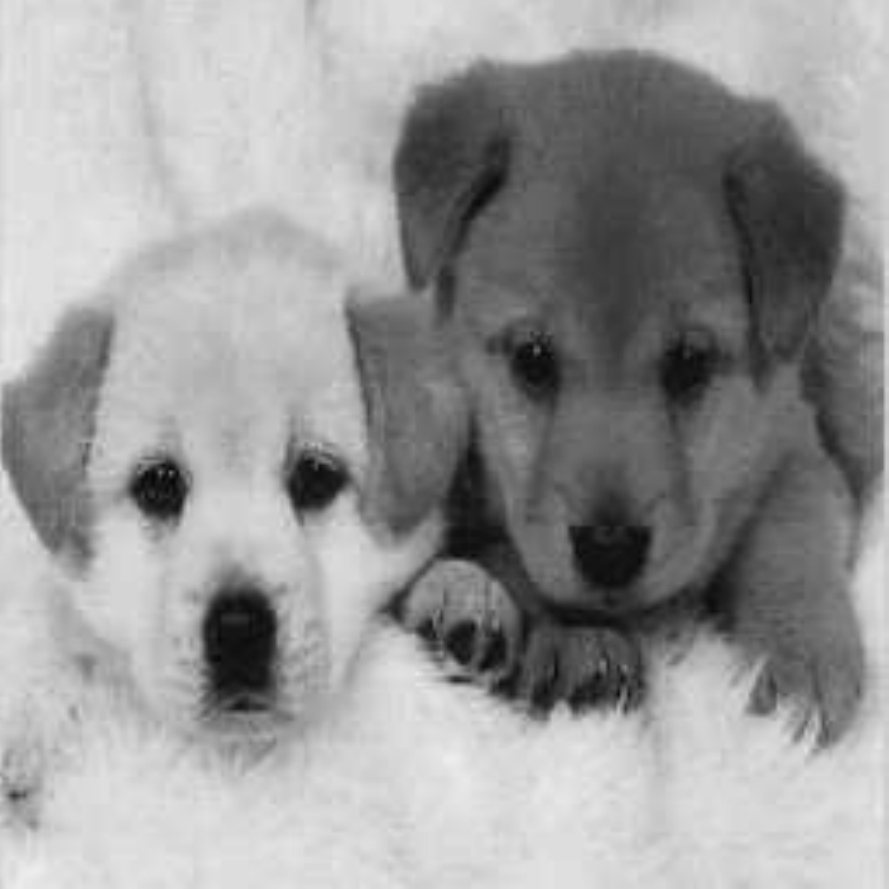}\\
Image \#2
\end{minipage}
\begin{minipage}{\sfigwidth}
\centering
\includegraphics[width=\textwidth]{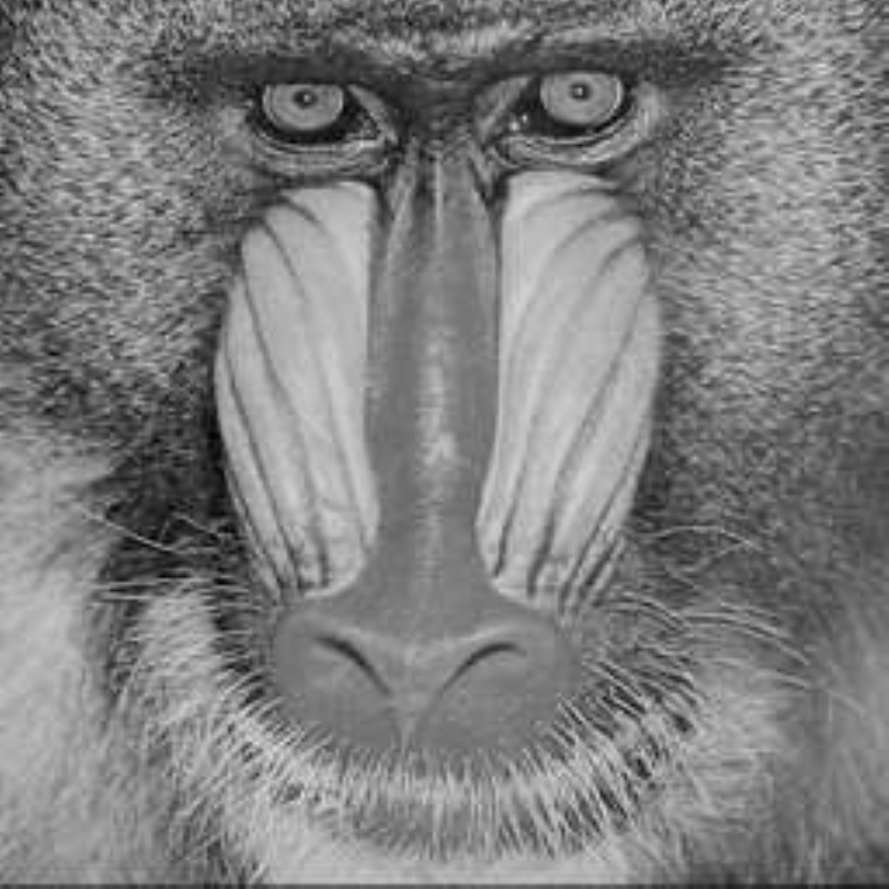}
Image \#3
\end{minipage}\\[\vfigskip]
\begin{minipage}{\sfigwidth}
\centering
\includegraphics[width=\textwidth]{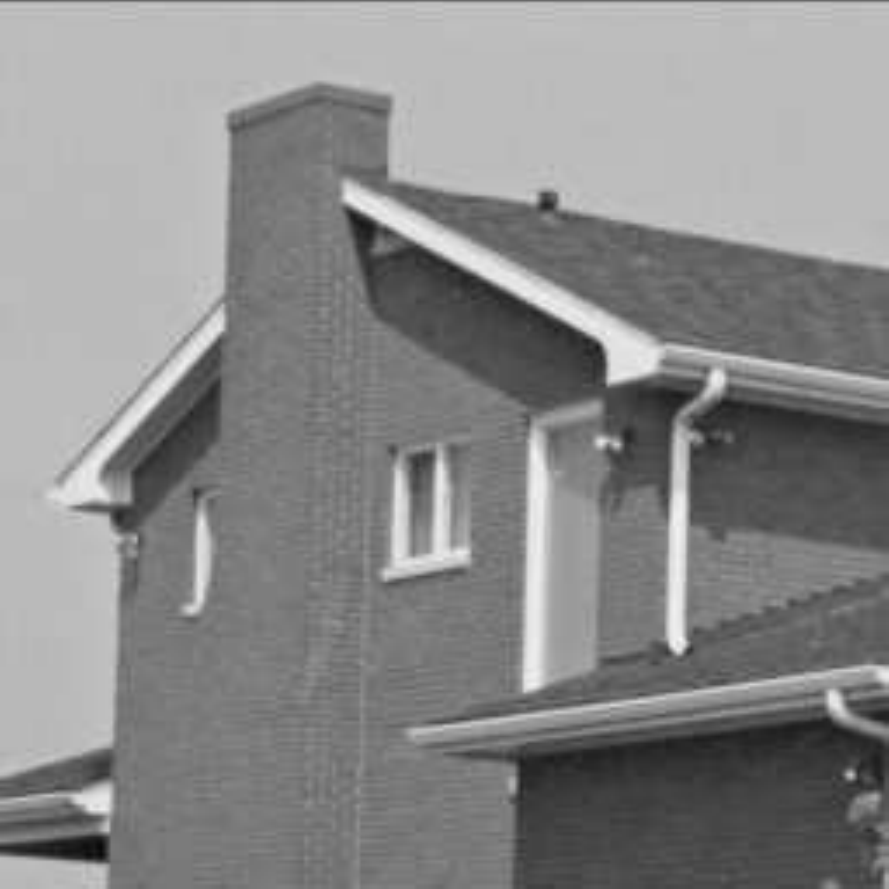}
Image \#4
\end{minipage}
\begin{minipage}{\sfigwidth}
\centering
\includegraphics[width=\textwidth]{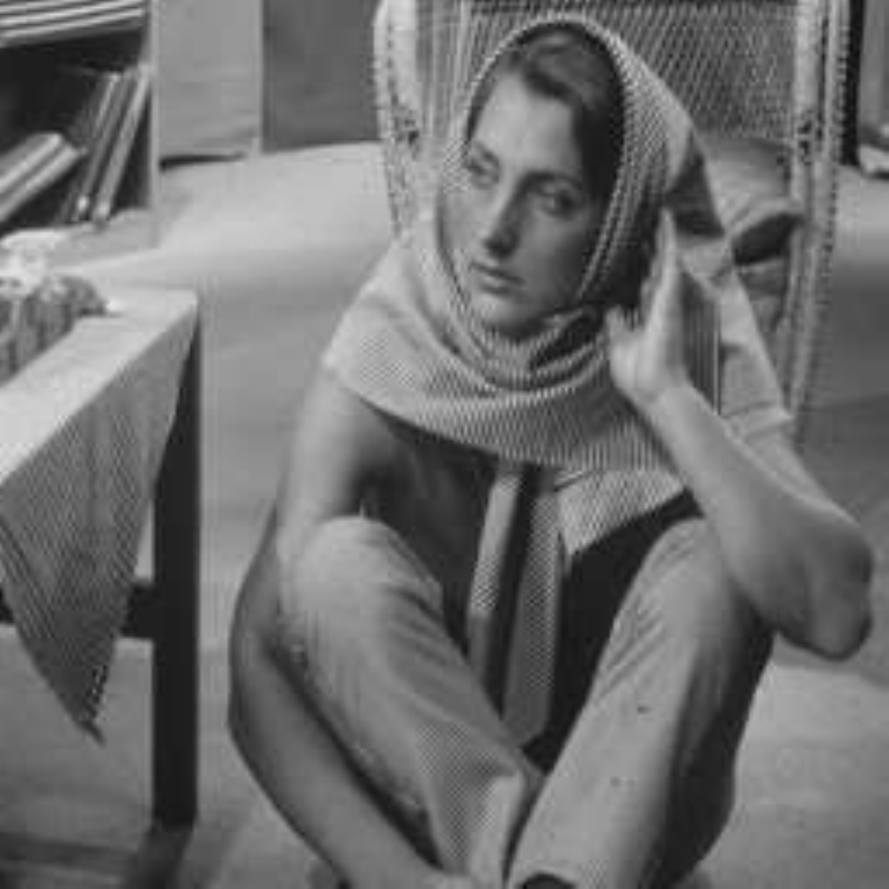}
Image \#5
\end{minipage}
\begin{minipage}{\sfigwidth}
\centering
\includegraphics[width=\textwidth]{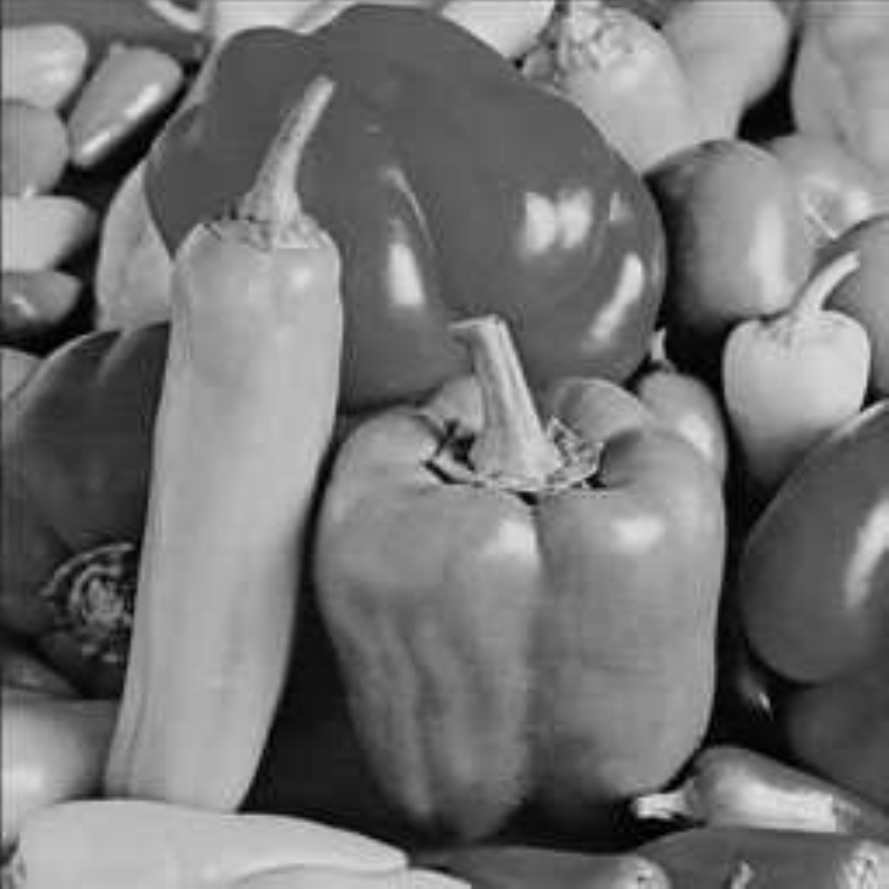}
Image \#6
\end{minipage}
\caption{The six $256\times 256$ test images used in the
experiments.} \label{figure:6TestImages}
\end{figure}

To verify the decryption performance of the above-discussed
known-plaintext attack to HCIE, some
experiments are performed using the six $256\times 256$ test
images with 256 gray scales shown in Fig.~\ref{figure:6TestImages}. Assume that the first $n=1\sim 5$ test
images are known to an attacker, the cipher-image of the last test
image is decrypted with the estimated permutation relationship matrices, to evaluate
the breaking performance.

In the experiments, three different configurations of HCIE are
used: $S_M=S_N=256$, $S_M=S_N=32$, $S_M=$ $S_N=16$. As mentioned
above, the configuration of $S_M=S_N=256$ corresponds to general
permutation-only image ciphers working in the spatial domain
(without using hierarchical structures). As shown in \cite[Sec.~4]{Li:Permutation:SPIC2008}, three
known plain-images are always enough to achieve a good breaking
performance, and an almost perfect breaking performance can
be achieved with four plain-images, where the public parameters are $\alpha=6$, $\beta=3$, $\gamma=3$ and
$no=9$.

\subsubsection{The experimental results with $S_M=S_N=32$}

\begin{figure}[!htb]
\centering
\begin{minipage}{\sfigwidth}
\centering
\includegraphics[width=\textwidth]{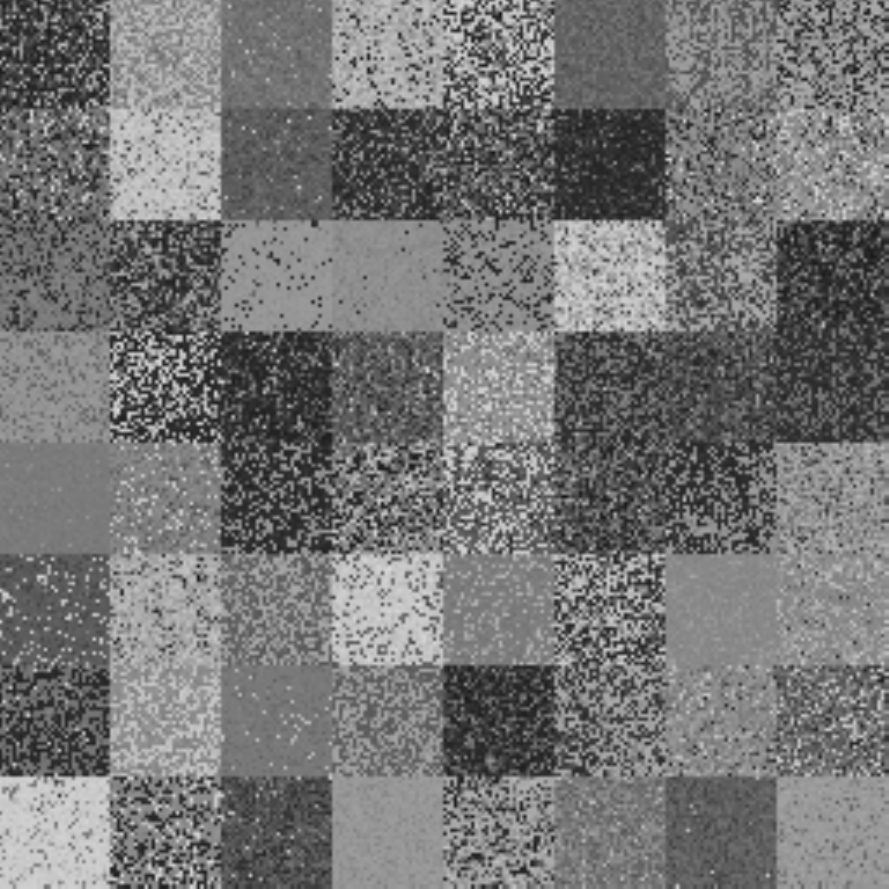}\\
Cipher-image \#1
\end{minipage}
\begin{minipage}{\sfigwidth}
\centering
\includegraphics[width=\textwidth]{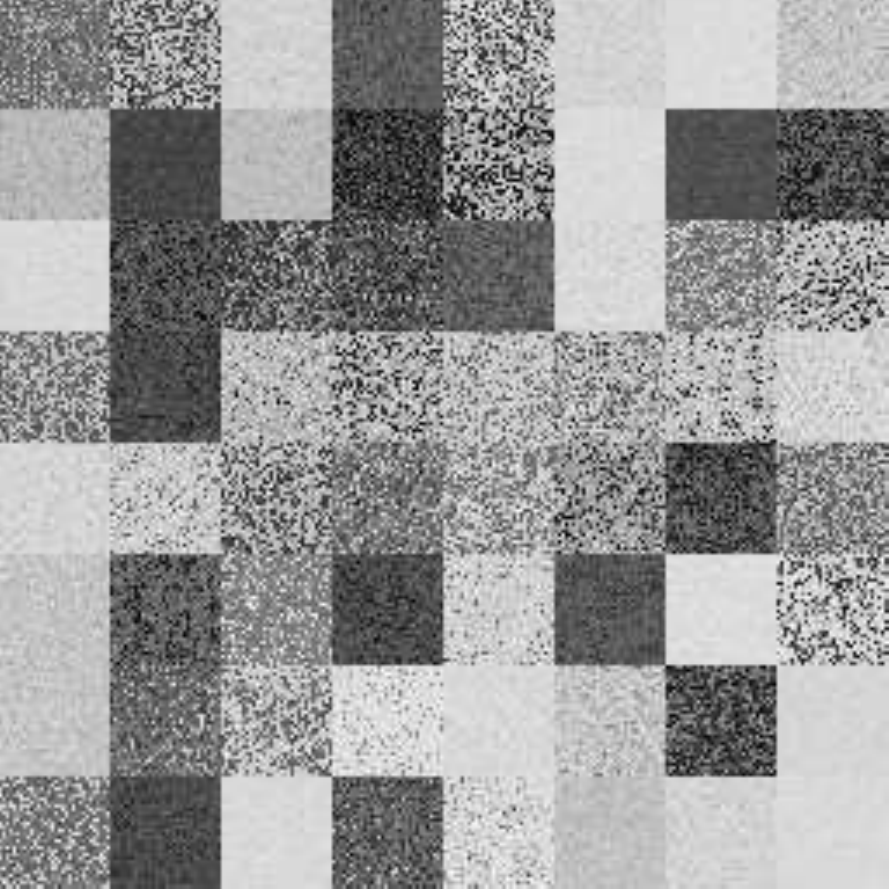}\\
Cipher-image \#2
\end{minipage}
\begin{minipage}{\sfigwidth}
\centering
\includegraphics[width=\textwidth]{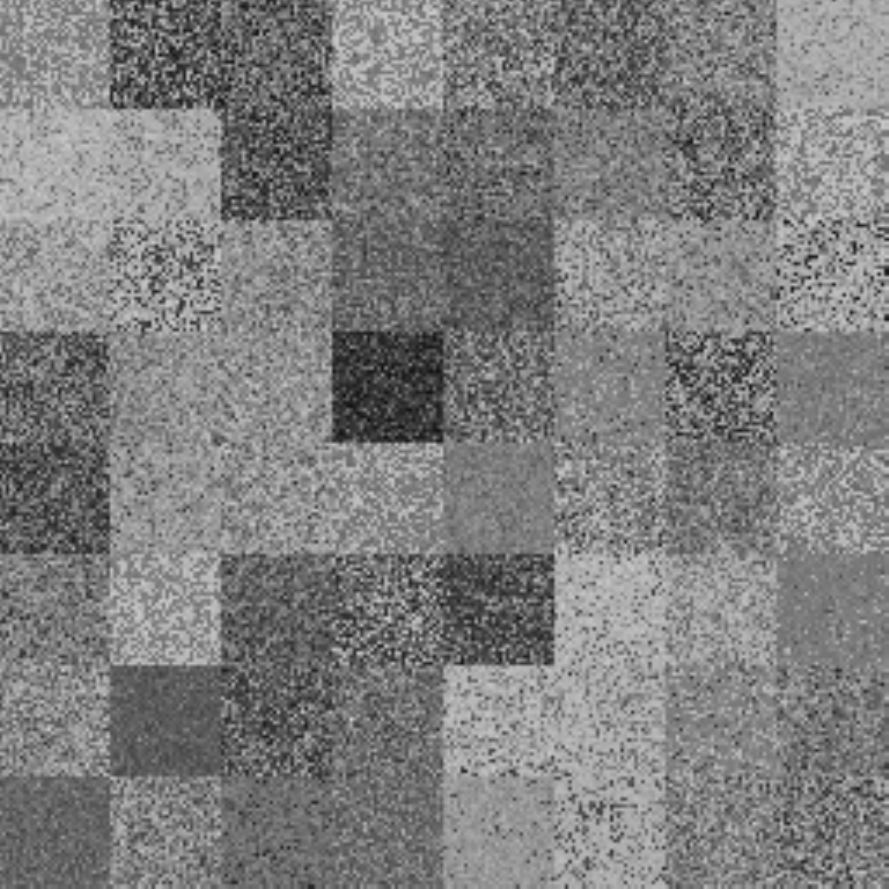}
Cipher-image \#3
\end{minipage}\\[\vfigskip]
\begin{minipage}{\sfigwidth}
\centering
\includegraphics[width=\textwidth]{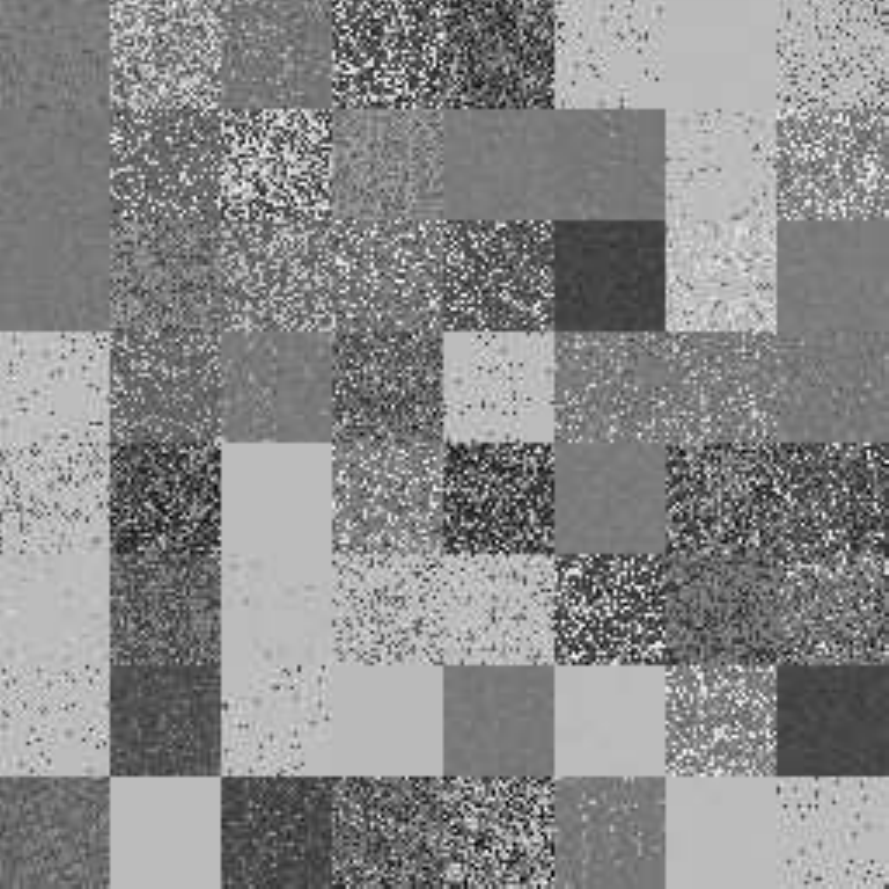}
Cipher-image \#4
\end{minipage}
\begin{minipage}{\sfigwidth}
\centering
\includegraphics[width=\textwidth]{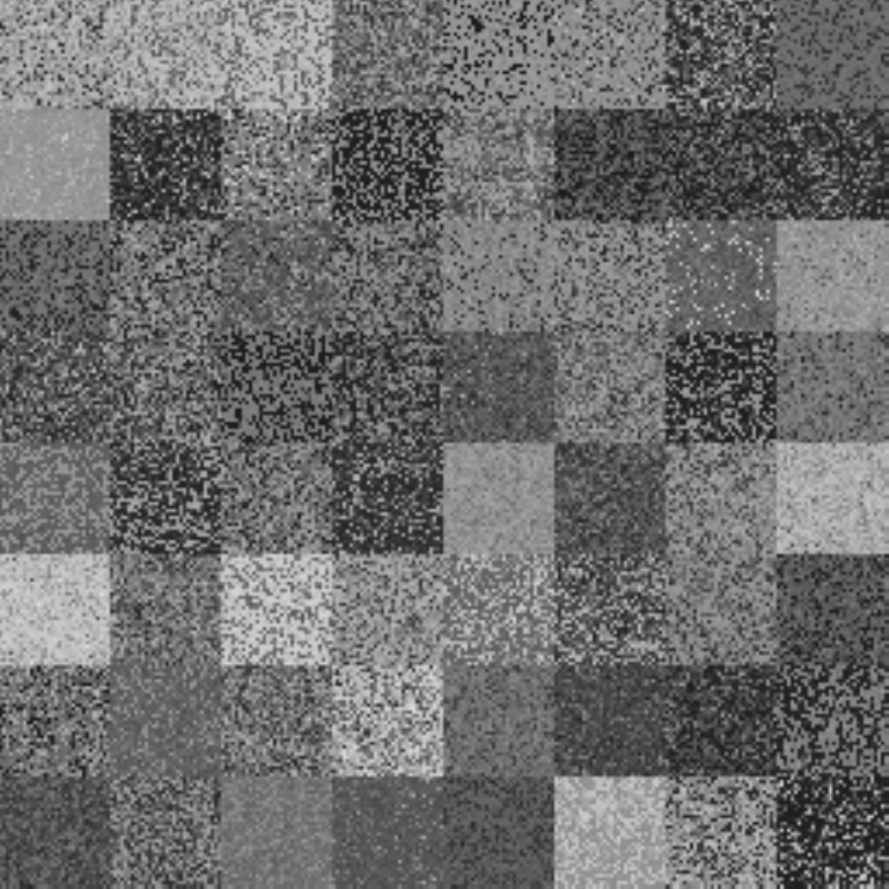}
Cipher-image \#5
\end{minipage}
\begin{minipage}{\sfigwidth}
\centering
\includegraphics[width=\textwidth]{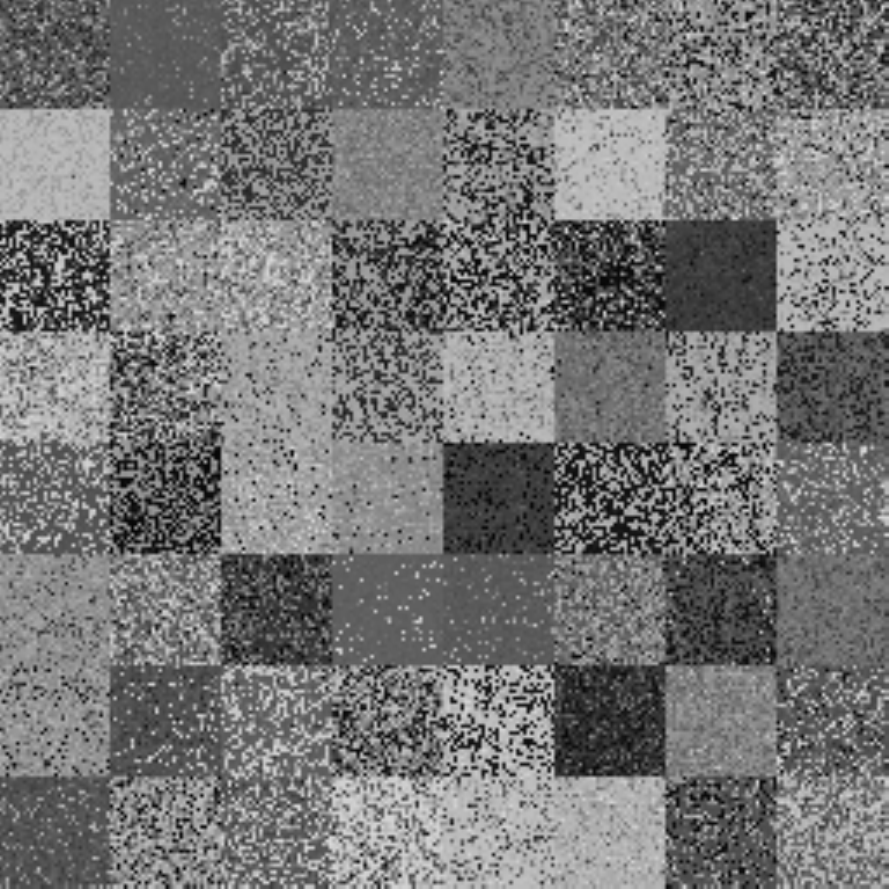}
Cipher-image \#6
\end{minipage}
\caption{The cipher-images of the six $256\times 256$ test images,
when $S_M=S_N=32$.} \label{figure:Experiment32}
\end{figure}

\begin{figure}[!htb]
\centering
\begin{minipage}{\sfigwidth}
\centering
\includegraphics[width=\textwidth]{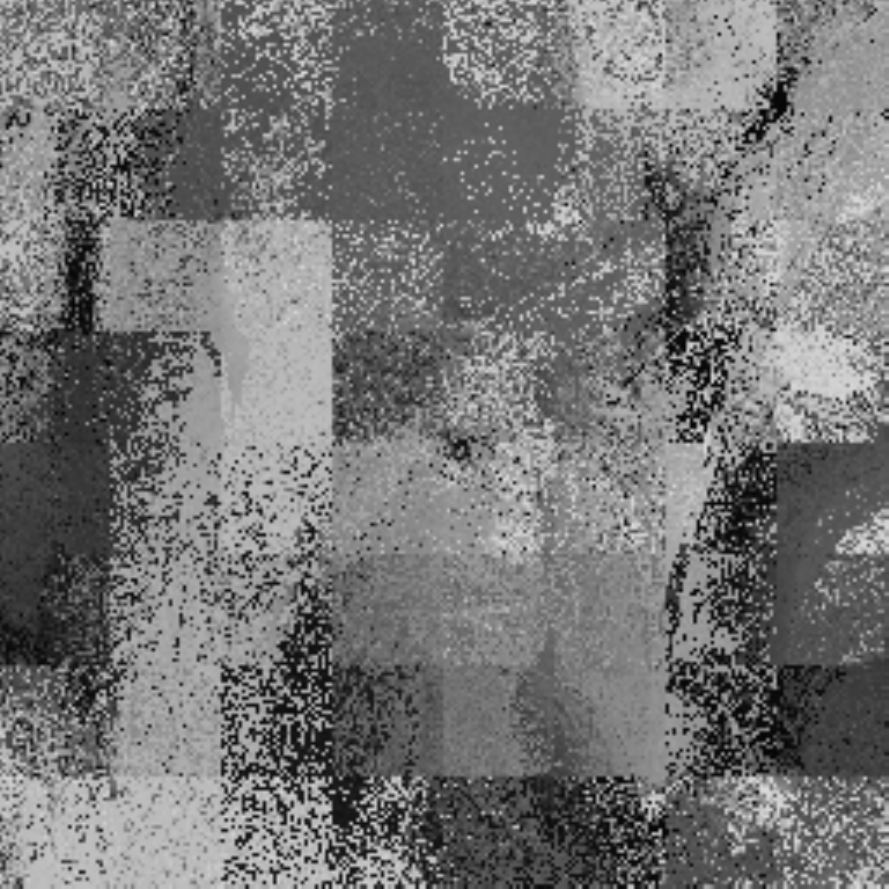}\\
$n=1$
\end{minipage}
\begin{minipage}{\sfigwidth}
\centering
\includegraphics[width=\textwidth]{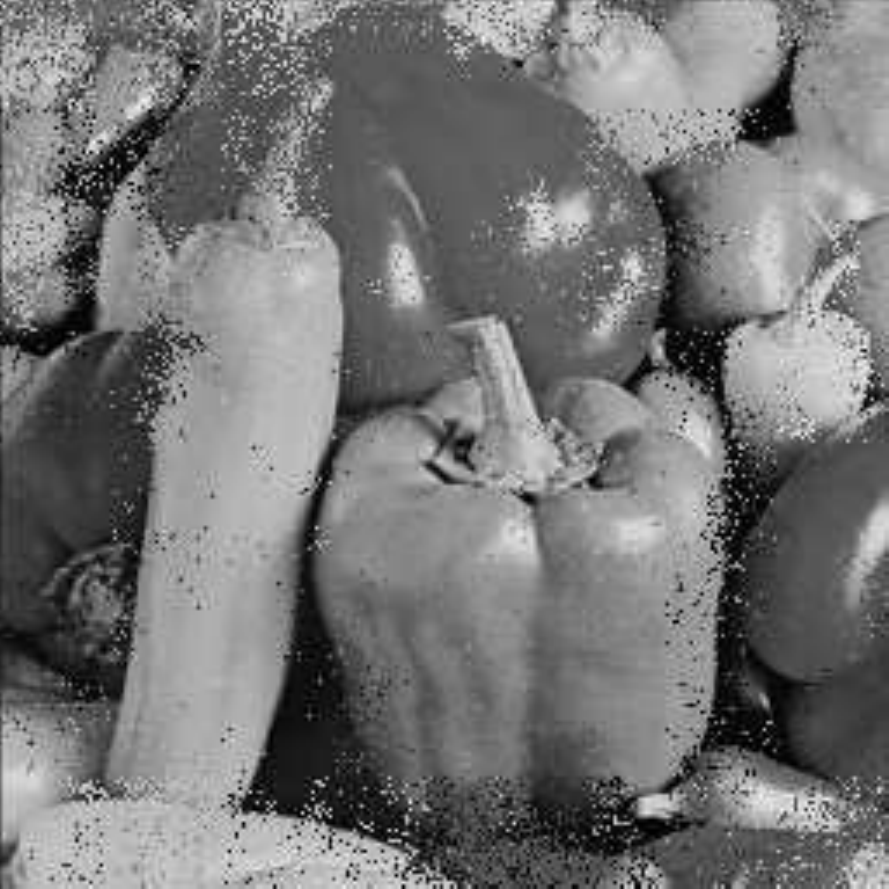}\\
$n=2$
\end{minipage}
\begin{minipage}{\sfigwidth}
\centering
\includegraphics[width=\textwidth]{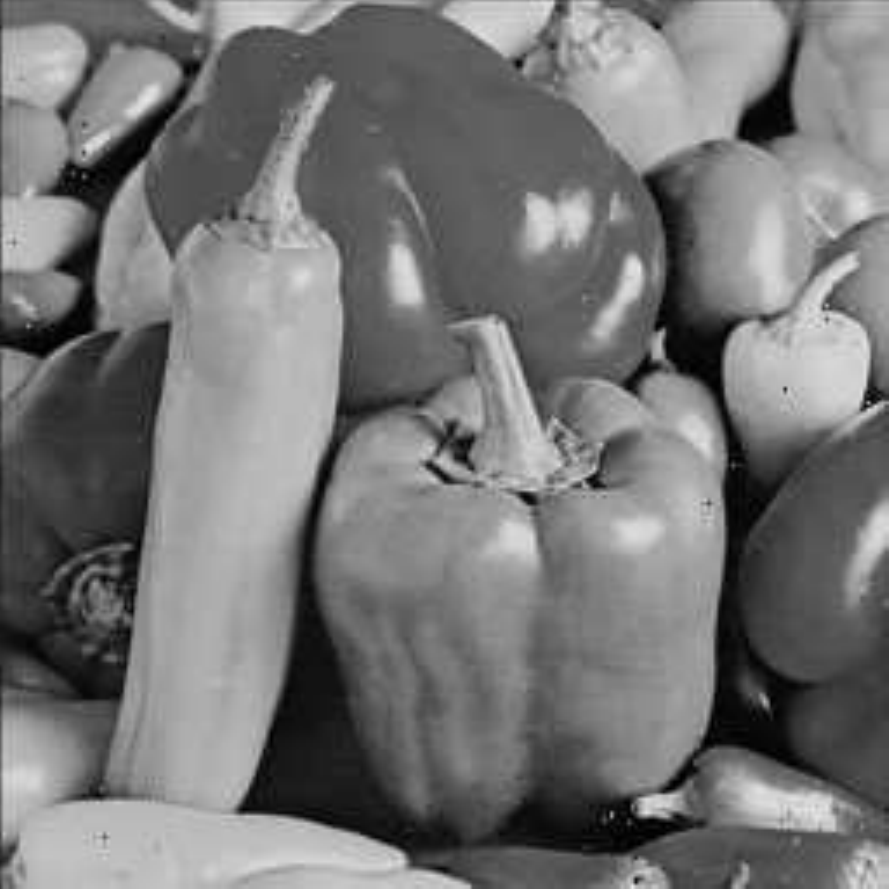}
$n=3$
\end{minipage}\\[\vfigskip]
\begin{minipage}{\sfigwidth}
\centering
\includegraphics[width=\textwidth]{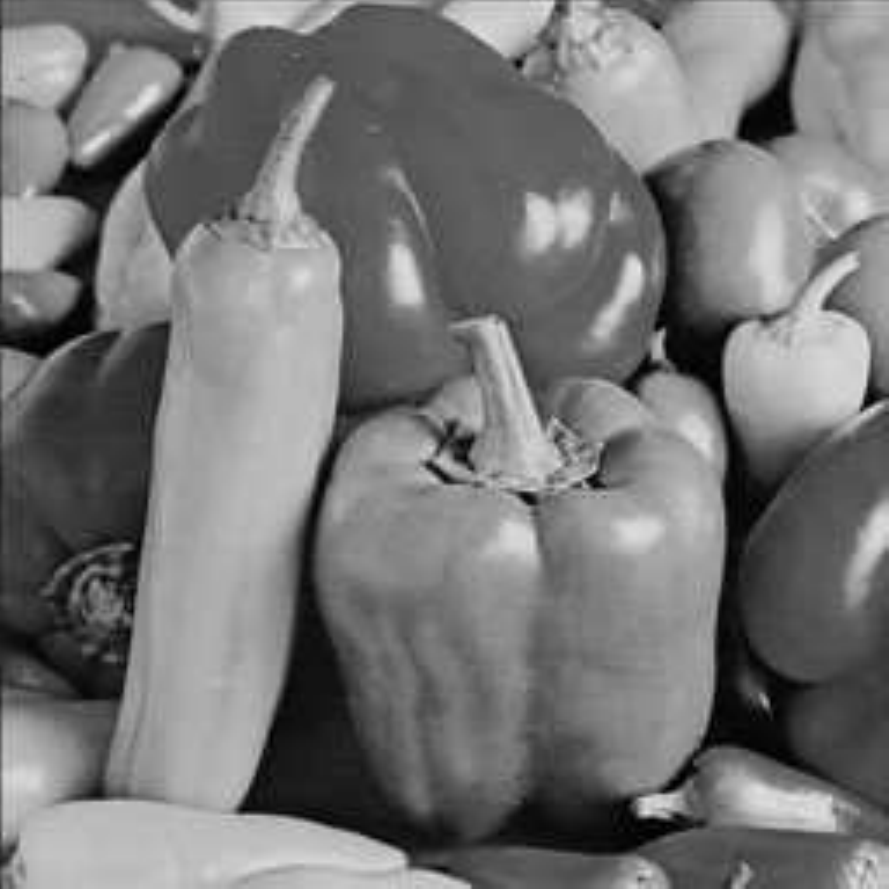}
$n=4$
\end{minipage}
\begin{minipage}{\sfigwidth}
\centering
\includegraphics[width=\textwidth]{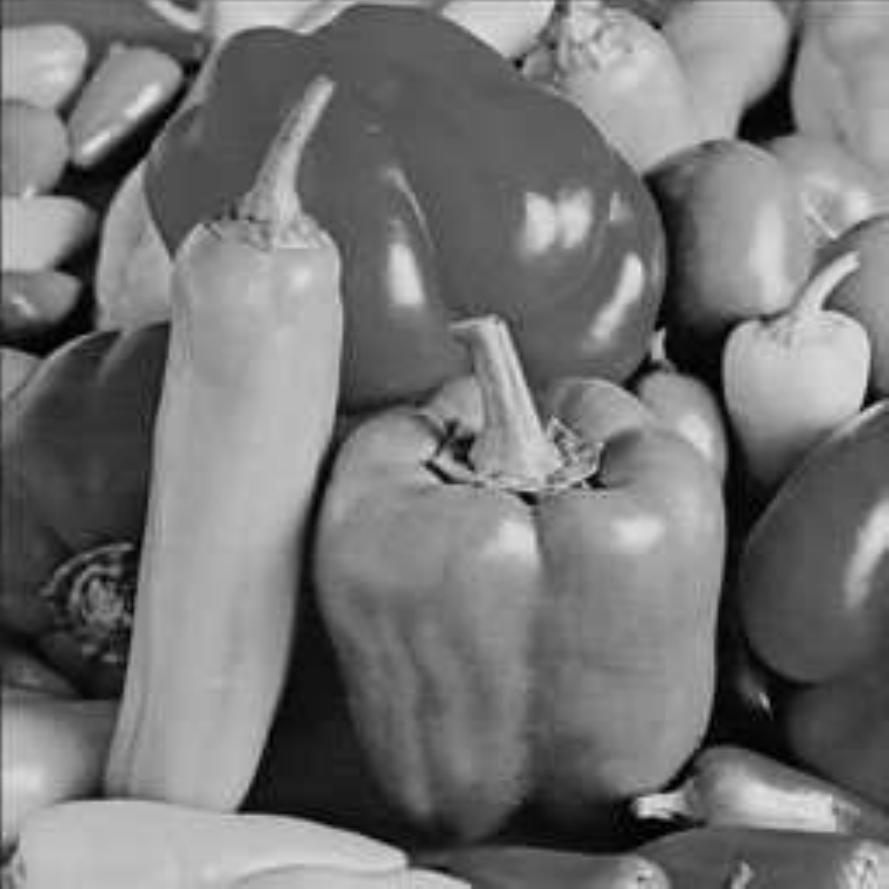}
$n=5$
\end{minipage}
\caption{The decrypted image of Cipher-Image \#6 when the first
$n$ test images are known to the attacker, when $S_M=S_N=32$.}
\label{figure:Decrypted32}
\end{figure}

The public parameters are set as follows: $\alpha=4$, $\beta=2$, $\gamma=1$ and
$no=2$. The cipher-images of the six test images are all shown in
Fig. \ref{figure:Experiment32}. When the first $n=1\sim 5$ test
images are known to the attacker, the obtained five decrypted images of the
sixth cipher-image are shown in Fig. \ref{figure:Decrypted32}. As
can be seen, one known plain-image cannot reveal much useful
visual information, but two is enough to obtain a good
performance.

\subsubsection{The experimental results with $S_M=S_N=16$}

\begin{figure}[!htb]
\centering
\begin{minipage}{\sfigwidth}
\centering
\includegraphics[width=\textwidth]{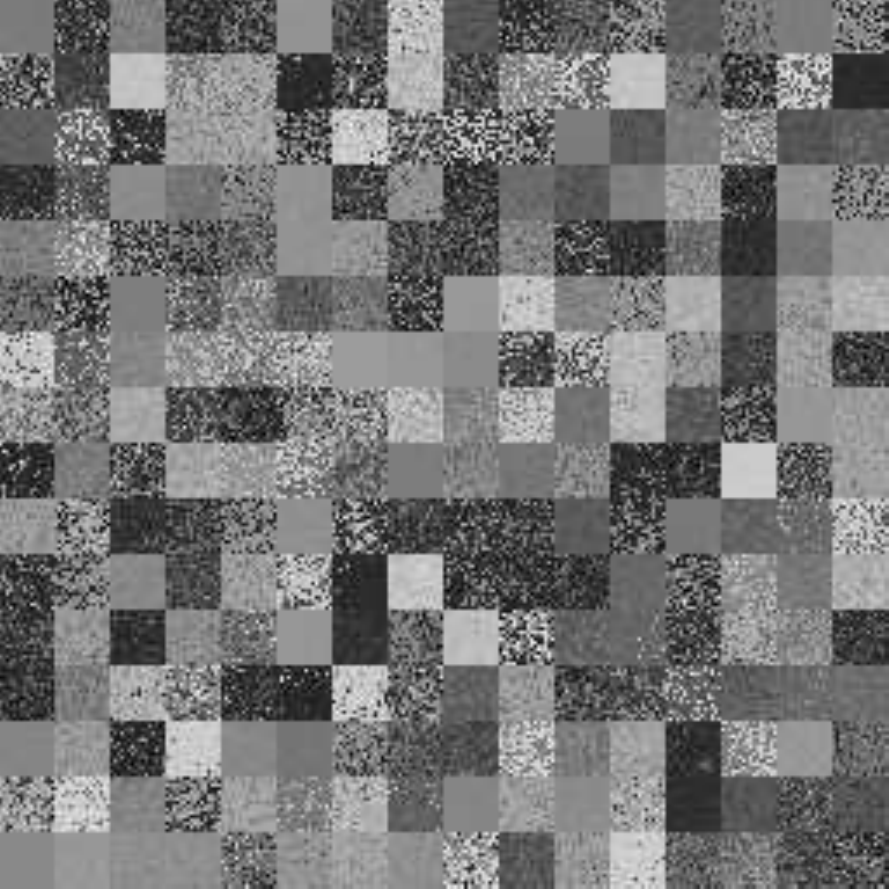}\\
Cipher-image \#1
\end{minipage}
\begin{minipage}{\sfigwidth}
\centering
\includegraphics[width=\textwidth]{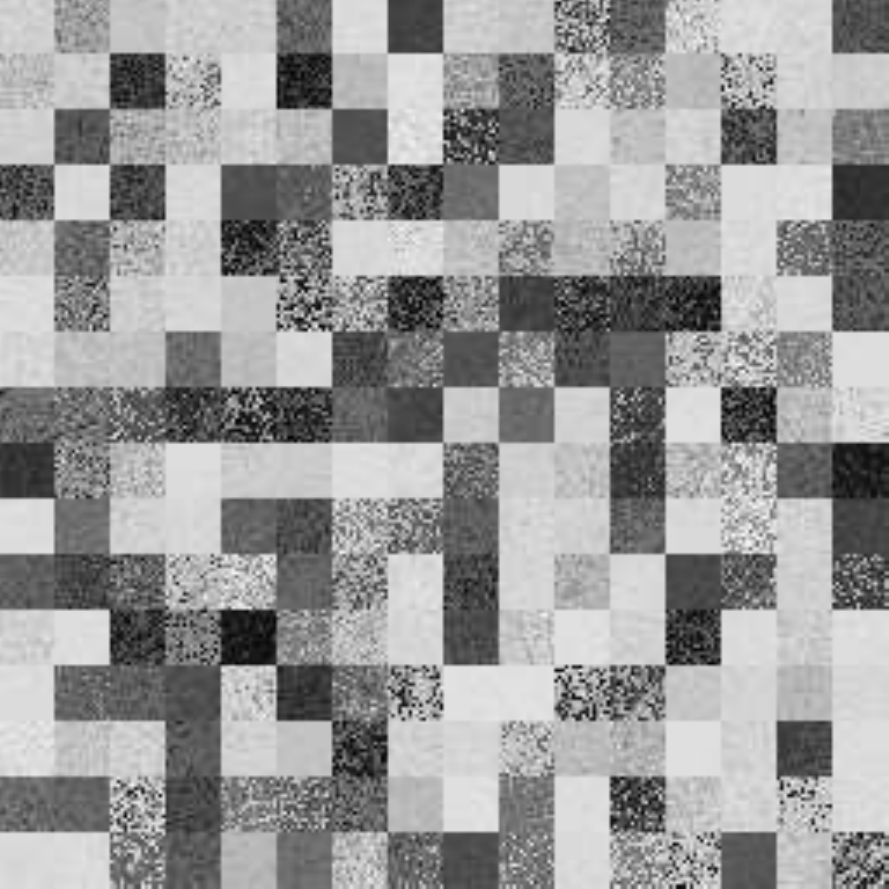}\\
Cipher-image \#2
\end{minipage}
\begin{minipage}{\sfigwidth}
\centering
\includegraphics[width=\textwidth]{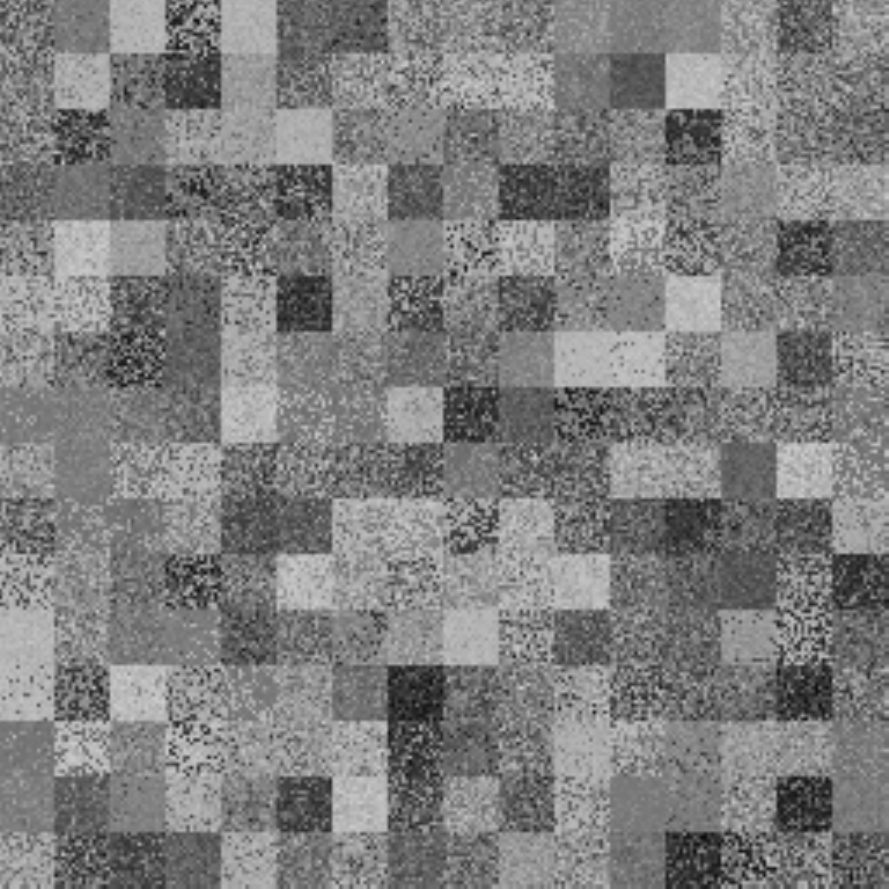}
Cipher-image \#3
\end{minipage}\\[\vfigskip]
\begin{minipage}{\sfigwidth}
\centering
\includegraphics[width=\textwidth]{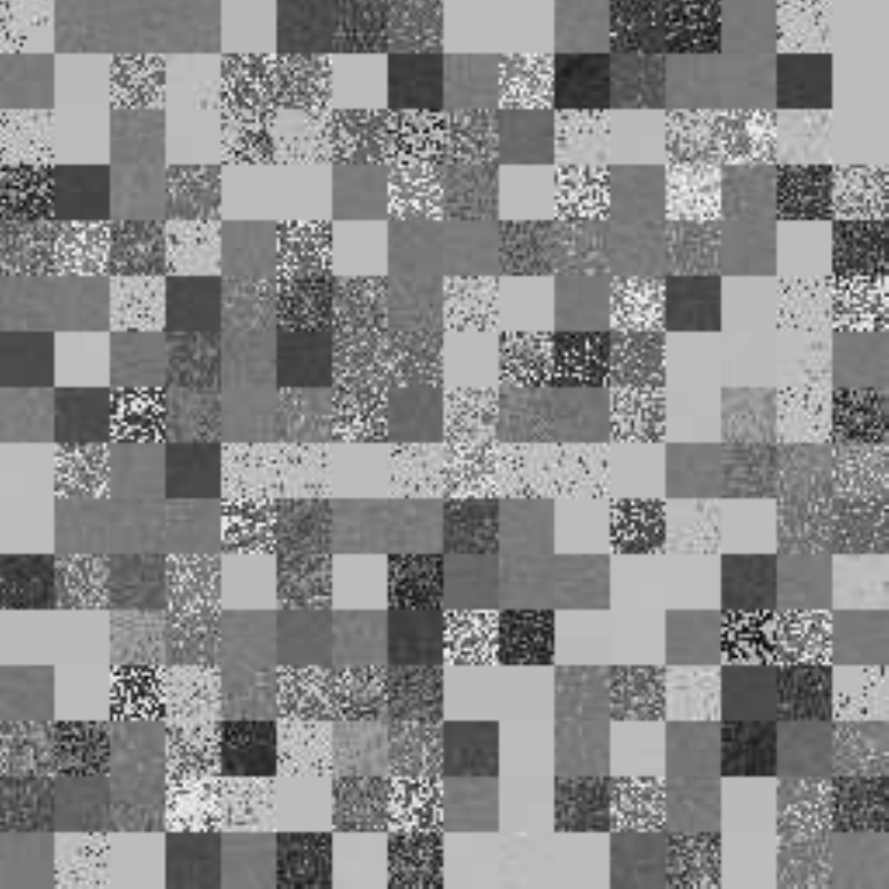}
Cipher-image \#4
\end{minipage}
\begin{minipage}{\sfigwidth}
\centering
\includegraphics[width=\textwidth]{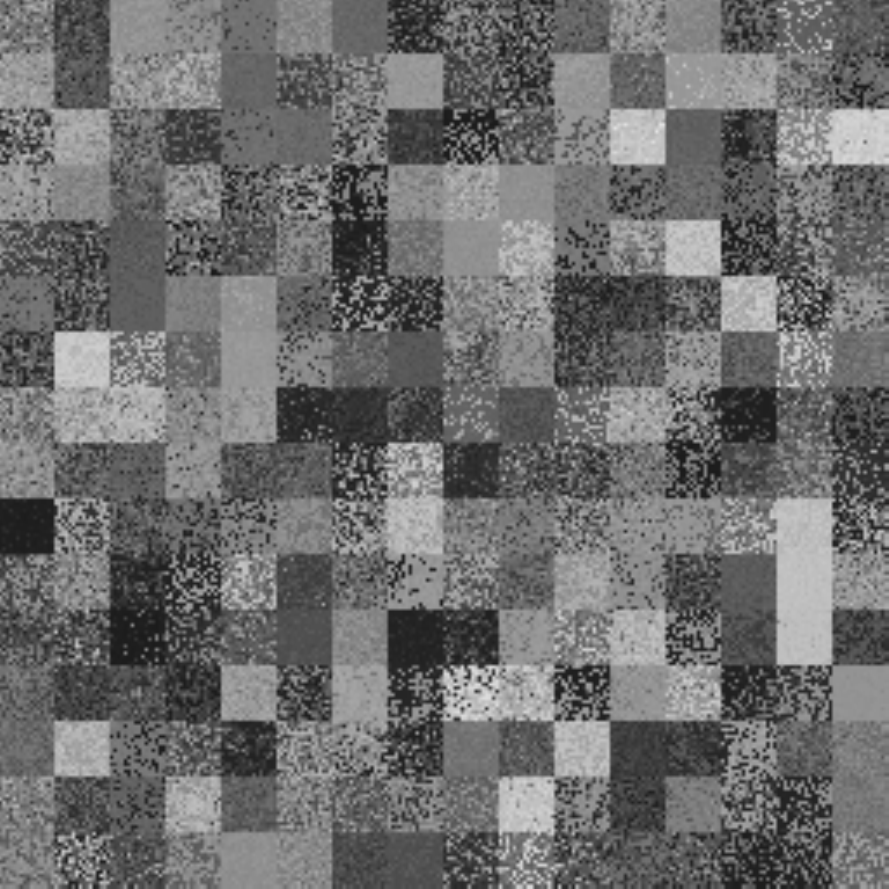}
Cipher-image \#5
\end{minipage}
\begin{minipage}{\sfigwidth}
\centering
\includegraphics[width=\textwidth]{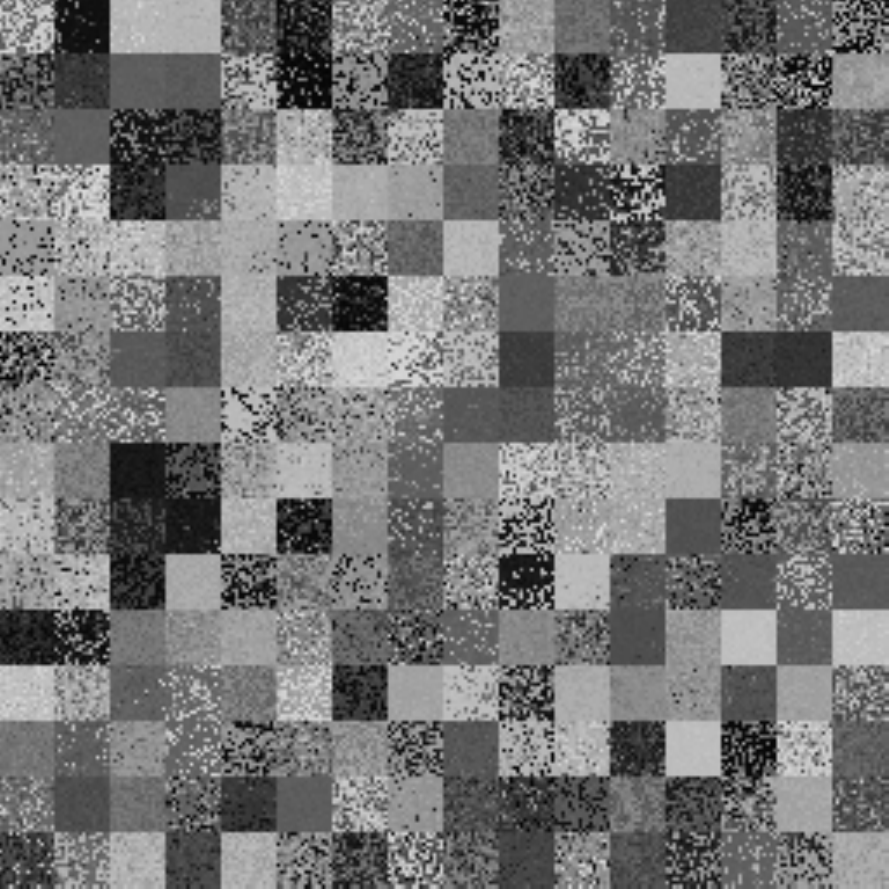}
Cipher-image \#6
\end{minipage}
\caption{The cipher-images of the six $256\times 256$ test images,
when $S_M=S_N=16$.} \label{figure:Experiment16}
\end{figure}

\begin{figure}[!htb]
\centering
\begin{minipage}{\sfigwidth}
\centering
\includegraphics[width=\textwidth]{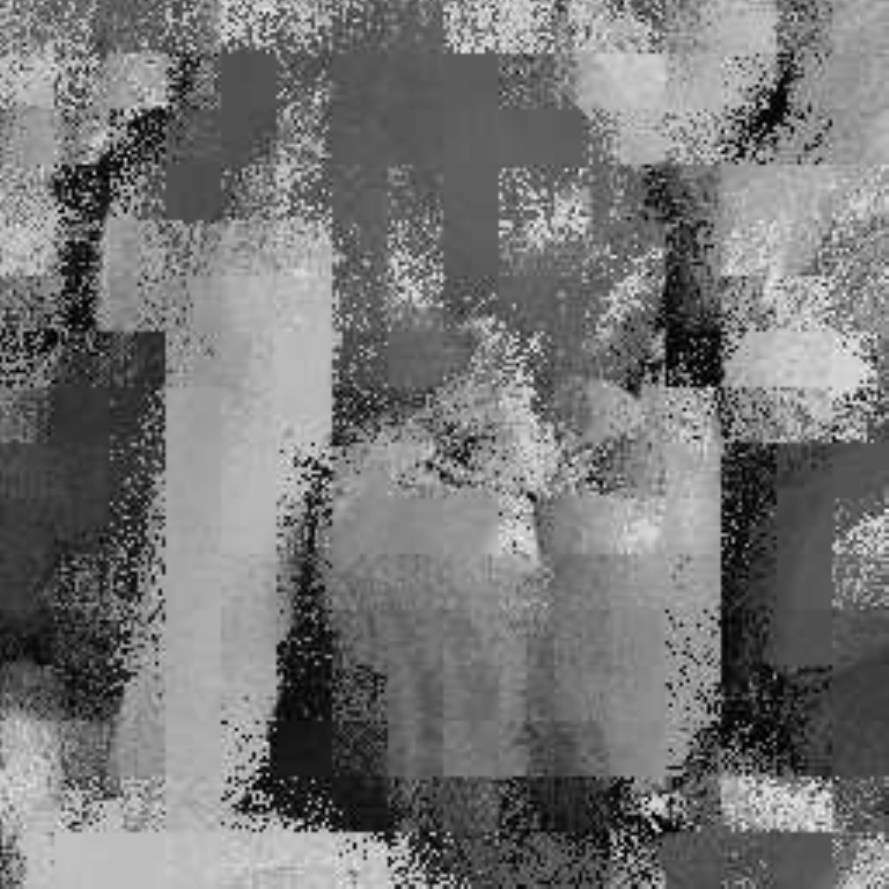}\\
$n=1$
\end{minipage}
\begin{minipage}{\sfigwidth}
\centering
\includegraphics[width=\textwidth]{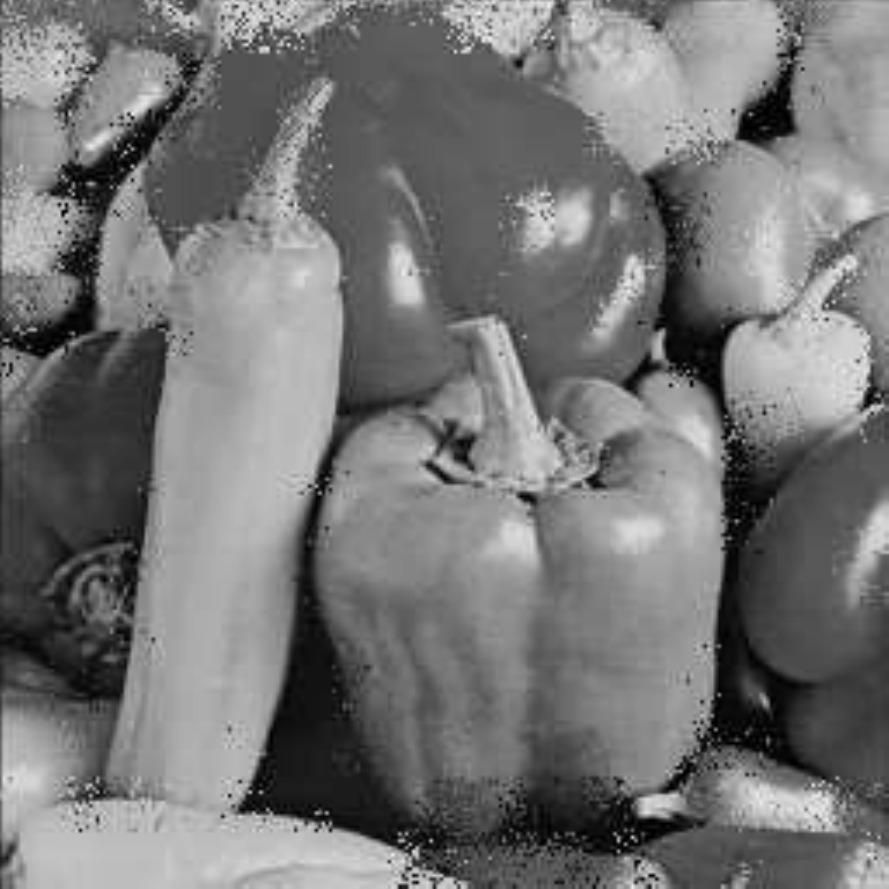}\\
$n=2$
\end{minipage}
\begin{minipage}{\sfigwidth}
\centering
\includegraphics[width=\textwidth]{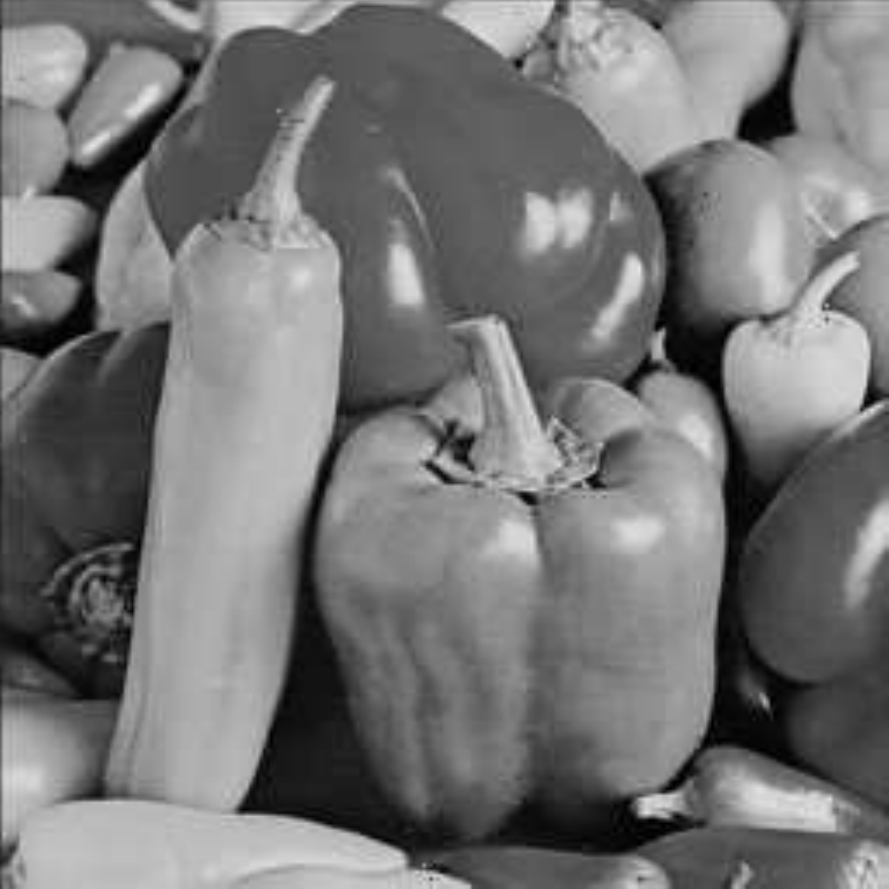}
$n=3$
\end{minipage}\\[\vfigskip]
\begin{minipage}{\sfigwidth}
\centering
\includegraphics[width=\textwidth]{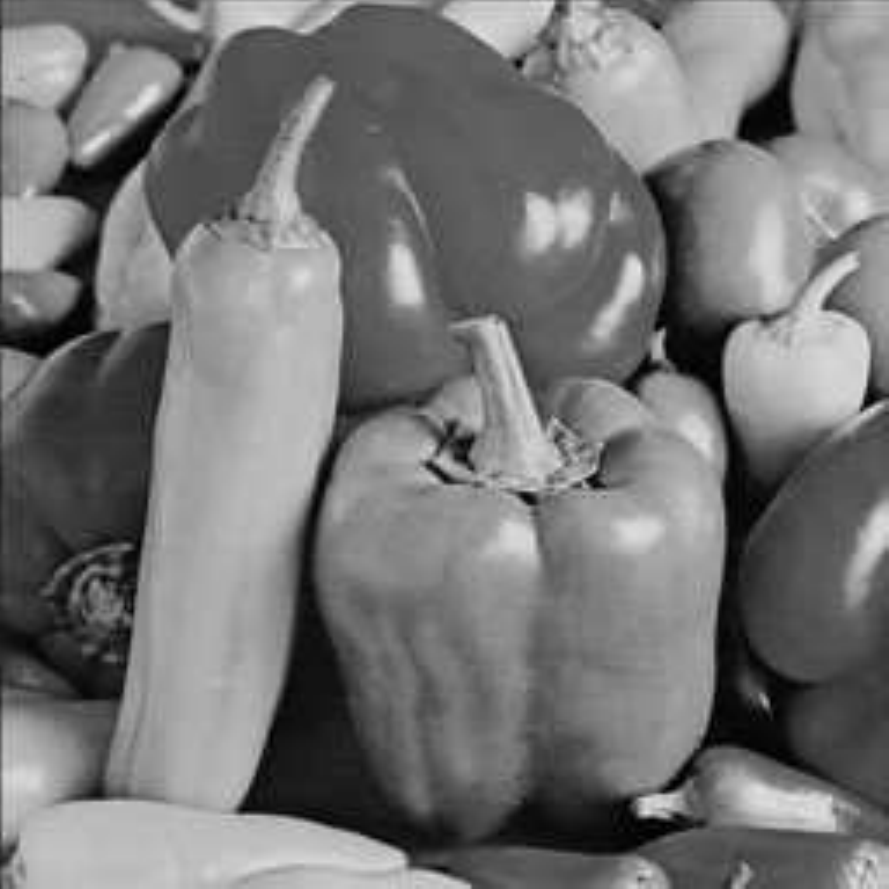}
$n=4$
\end{minipage}
\begin{minipage}{\sfigwidth}
\centering
\includegraphics[width=\textwidth]{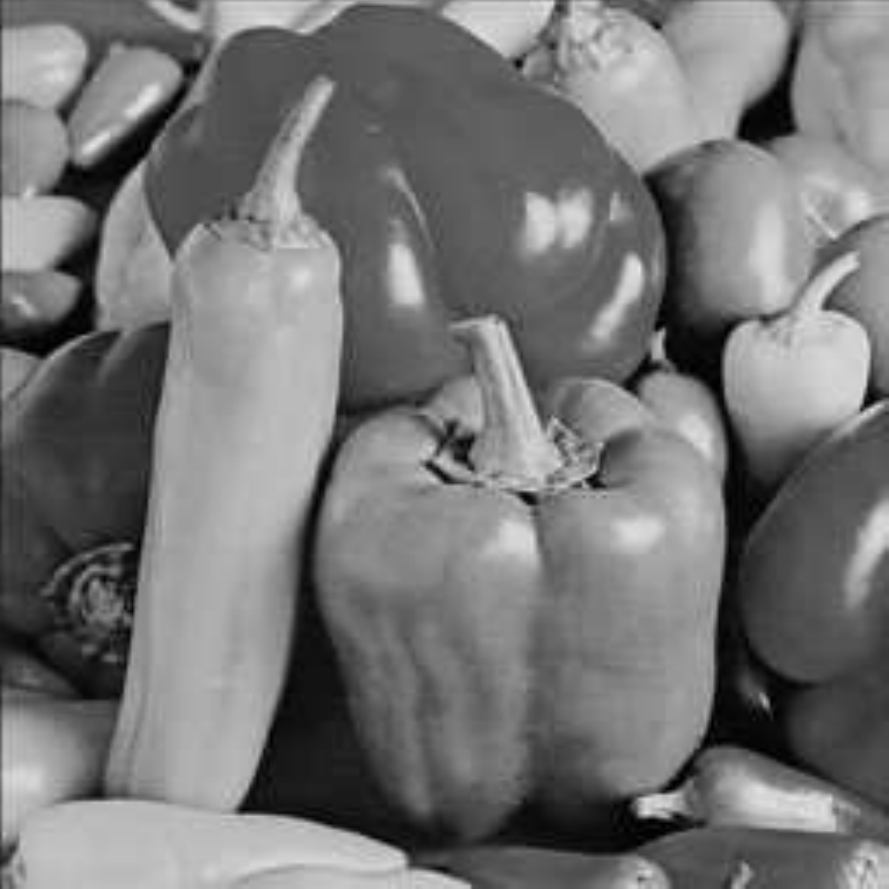}
$n=5$
\end{minipage}
\caption{The decrypted images of Cipher-Image \#6 when the first
$n$ test images are known to an attacker, when $S_M=S_N=16$.}
\label{figure:Decrypted16}
\end{figure}

The public parameters are $\alpha=4$, $\beta=2$, $\gamma=1$ and
$no=2$. The cipher-images of the six test images are all shown in
Fig. \ref{figure:Experiment16}. When the first $n=1\sim 5$ test
images are known to the attacker, the five decrypt images of the
sixth cipher-image are shown in Fig. \ref{figure:Decrypted16}. As
can be seen, even one known plain-image can reveal a rough view of
the plain-image, and two is enough to obtain a nearly-perfect
recovery.

\begin{figure}[!htb]
\centering
\begin{overpic}[width=2\figwidth]{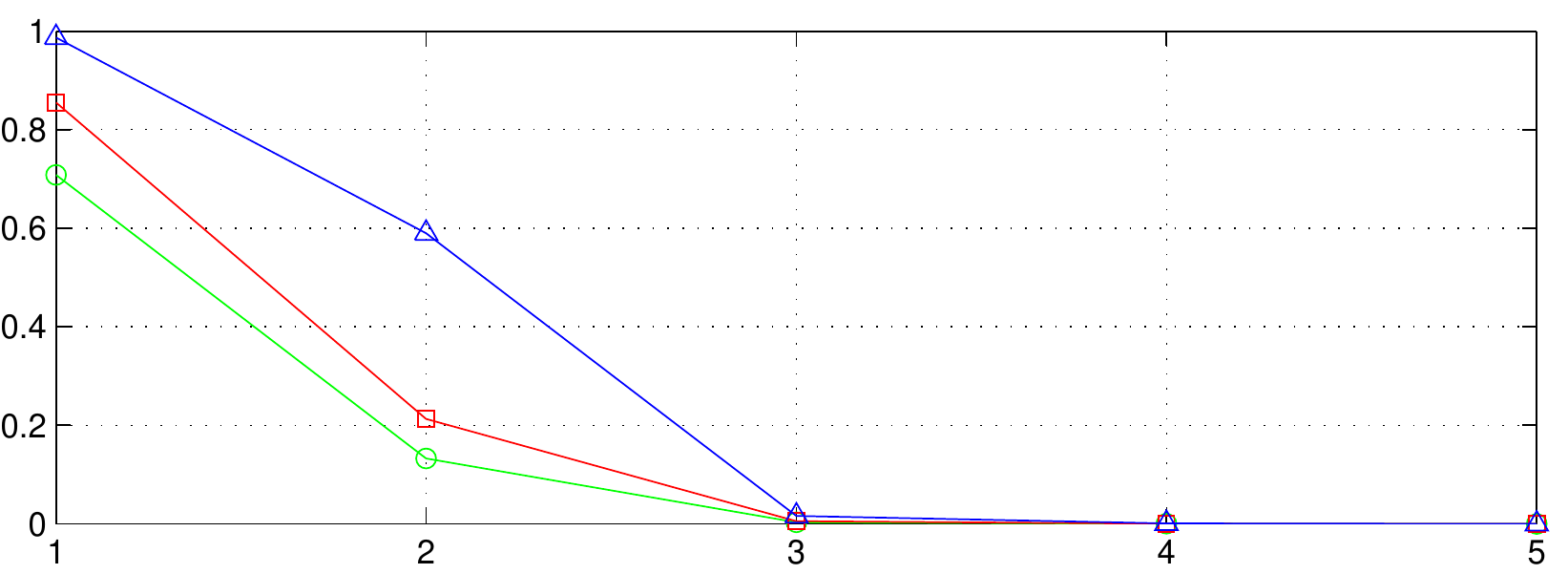}
    \put(50,-3){$n$}
\end{overpic}\\[1em]
a) decryption error ratio\\
\begin{overpic}[width=2\figwidth]{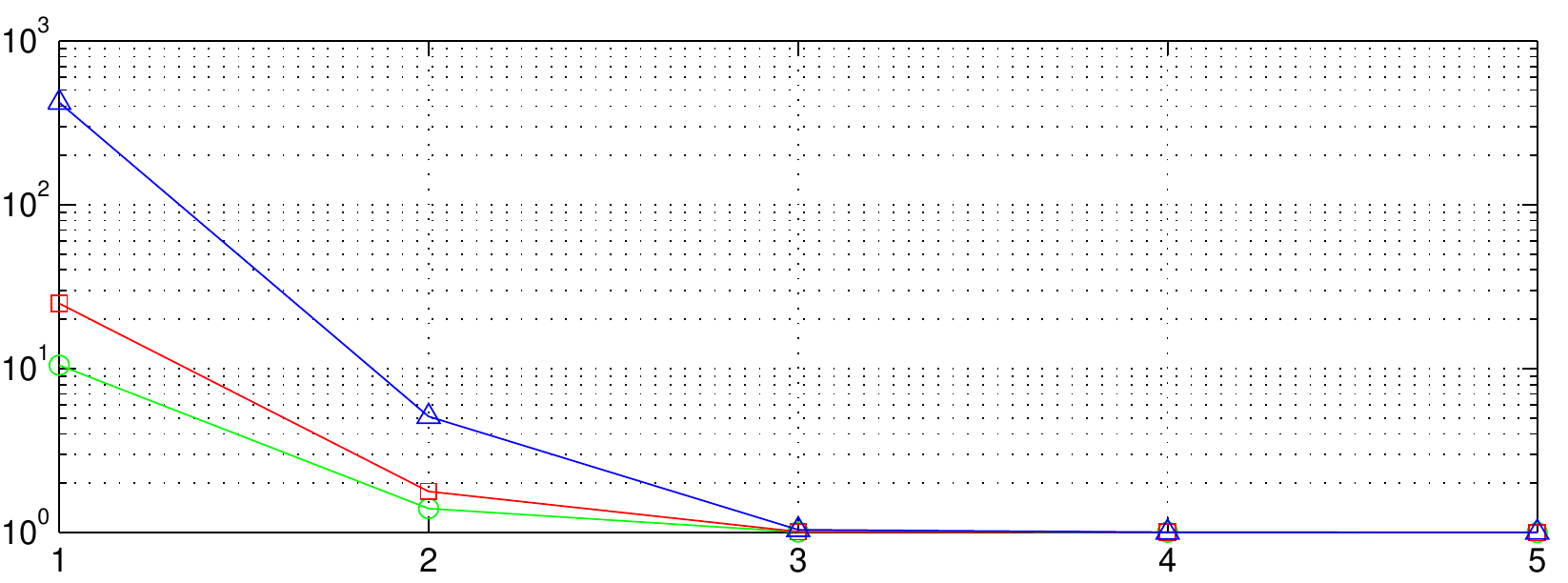}
    \put(50,-3){$n$}
\end{overpic}\\[1em]
b) the average cardinality $\overline{\#(\widehat{w}(i,j))}$\\
(Legend: \textcolor{blue}{$\bigtriangleup$} -- $S_M=S_N=256$,
\textcolor{red}{$\Box$} -- $S_M=S_N=32$,
\textcolor{green}{\footnotesize$\bigcirc$} -- $S_M=S_N=16$)
\caption{A performance comparison of the known-plaintext attack to
HCIE.}\label{figure:Comparison}
\end{figure}

Now, let us give a performance comparison of the
known-plaintext attack to HCIE with the above three different
configurations. Figure \ref{figure:Comparison}a) shows the
quantitative relationship between the number of known plain-images and
the decryption quality (represented by the decryption error
ratio). It can be seen that three known plain-images are enough
for all three configurations to achieve an acceptable breaking
performance, and two can reveal quite a lot of pixels (which means
that most significant visual information is revealed). Also, it is
found that the breaking performance is dependent on the
configuration: when $S_M=S_N=16$, the best performance is
achieved, which coincides with the previous analysis.

Figure \ref{figure:Comparison}b) shows the average cardinality of
the elements in $\widehat{\bm{W}}$, which is an indicator of the
probability of getting correct permutation elements in
$\widetilde{\bm{W}}$ and an indicator of the time complexity as
analyzed above. Comparing Figures \ref{figure:Comparison}a) and
\ref{figure:Comparison}b), one can see that the occurrence
probability of decryption errors has a good correspondence with
the average cardinality, where the correctness of the uniquely-determined permutation relationship matrix
was obtained by some chosen plain-images and the corresponding cipher-images.

From the above comparison, it is concluded that the security of HCIE
with a hierarchical structure is even weaker than the security of
general permutation-only image encryption algorithms without hierarchical
structures: when $S_M=S_N=32$ and $S_M=S_N=16$, two known
plain-images are enough to achieve an acceptable breaking
performance; while when $S_M=S_N=256$, the breaking performance
with two known plain-images is not satisfactory, thus three
plain-images are needed to achieve an acceptable performance.
Overall, from the viewpoint of security against
known/chosen-plaintext attacks, the hierarchical idea proposed in
HCIE has no technical merits.

\subsection{Chosen-plaintext attack}

To discover an equivalent secret key of a common permutation-only image encryption scheme under the scenario of chosen-plaintext attack, one can construct a ``composition plain-text", whose every element is different from each other \cite[Sec.~5.1]{Li:TDCEA:JASP2005}. To satisfy the requirement, the number of the bit planes of the special chosen-text should be not smaller than $\lceil \log_2(M\cdot N)\rceil$.
So, the number of required chosen-images is
\begin{equation*}
n=\lceil \lceil\log_2(M\cdot N)\rceil/\lceil\log_2(T)\rceil \rceil.
\end{equation*}
In general, $T$ is a power of 2 in the digital domain, hence $\log_2(T)$ is an integer. To this end, one has
\begin{equation}
n= \lceil\log_2(M\cdot N)/\log_2(T)\rceil =\lceil \log_T(M\cdot N)\rceil
\label{eq:numofchosenplaintext}
\end{equation}
by referring to \cite[Theorem 3.10]{Knuth:Concrete:1989}.

Similarly to the known-plaintext
attack, the use of a hierarchical structure in HCIE can also make
the construction of chosen plain-images easier. Accordingly, an
attacker can work hierarchically to construct $n$ chosen
plain-images, $f_1,\cdots,f_n$, as follows:
\begin{itemize}
\item \textit{high-level}:
$\overline{P}_{f_1}\sim\overline{P}_{f_n}$, which are defined in
Eq. (\ref{equation:PlainBlock}), compose an orthogonal image set;

\item \textit{low-level}: $\forall(i,j)$, $P_{f_1}(i,j)\sim
P_{f_n}(i,j)$ compose an orthogonal image set.
\end{itemize}
In this case, the minimal number of required chosen plain-images
becomes
\begin{eqnarray}
n & = & \max\left(\left\lceil\log_T(S_M\cdot
S_N)\right\rceil,\left\lceil\log_T\left(K\right)\right\rceil\right)\nonumber\\
& \leq &
\left\lceil\log_T(M\cdot N)\right\rceil,\label{equation:HCIE_plaintext_number}
\end{eqnarray}
where $K=\frac{M}{S_M}\cdot \frac{N}{S_N}$.
The above equality holds if and only if the hierarchical
encryption structure is disabled, i.e., when $K\in\{1, M\cdot N\}$.

As the $\left(1+\frac{M}{S_M}\cdot\frac{N}{S_N}\right)$ permutation relationship matrices of HCIE
are uniquely determined by the bit sequence $\{b(i)\}_{i=0}^{L_b-1}$ and the public parameters, one may recover reversely some consecutive bits of $\{b(i)\}_{i=0}^{L_b-1}$ \cite[Sec.~3.3.6]{Lcq:MCS:JSS10}. Furthermore, one can derive
the secret key $x(0)$ and $\mu$ following the approach given in \cite[Sec.3.3.2]{Li:RCES:JSS2008}.

\section{Conclusion}

Specific security performance of a typical permutation-only encryption algorithm, called HCIE, against ciphertext-only attack and known/chosen-plaintext attacks has been studied in detail. It is found that the capability of HCIE against the former attack was over-estimated much and hierarchical permutation-only image encryption algorithms such as HCIE are less secure than normal permutation-only ones without using hierarchical encryption structures. This work effectively demonstrates that the size of the real permutation domain of a permutation algorithm should be as large as possible in order to reach the best performance. As permutation operation alone cannot provide high level of security, it should be combined with other value substitution functions.

%\iffalse
\section*{Acknowledgement}

This research was supported by the Distinguished Young Scholar Program of the Hunan Provincial Natural Science Foundation of China (No.~2015JJ1013).
Some parts of Sec.~3 were completed with the help of \href{www.hooklee.com}{Dr.~Shujun Li}, from Surrey University, UK.
%\fi

\bibliographystyle{elsarticle-num}
%\nocite*
\bibliography{HCIE2}
\end{document}